\documentclass[11pt,a4paper]{article}
\usepackage{cite}
\usepackage{graphicx}
\usepackage{amssymb}
\usepackage{amsmath}
\usepackage{amsfonts}
\usepackage{dsfont}
\usepackage{mathtools}
\usepackage{slashed}
\usepackage{multirow}
\usepackage{rotating}
\usepackage{bbold,amsfonts}
\usepackage[utf8]{inputenc}
\usepackage{bm}
\usepackage{xcolor}
\usepackage{float}
\usepackage{braket}
\usepackage[height=8.8in,width=6.45in]{geometry}
\usepackage[font=small,labelfont=bf]{caption}
\usepackage[hidelinks]{hyperref}
\numberwithin{equation}{section}

\def\sp#1#2{\big[^{#1}_{#2}\big]}
\newcommand{\beq}{\begin{equation}}
\newcommand{\eeq}{\end{equation}}

\newcommand{\comm}[2]{\left[#1,#2\right]}

\DeclareMathOperator{\tr}{tr}

\newcommand{\one}{\mathbb{1}}
\newcommand{\ii}{\mathrm{i}}
\newcommand{\overbar}[1]{\mkern 1.5mu\overline{\mkern-1.5mu#1\mkern-1.5mu}
\mkern 1.5mu}
\makeatletter
\newcommand*{\letterdef@}{}
\newcommand*{\letterdef}[3]{%
	\def\letterdef@##1{\expandafter\newcommand\csname #1\endcsname{#2{##1}}}%
	\@tfor\@tempa :=#3\do{\expandafter\letterdef@\expandafter{\@tempa}}}
\makeatother
\letterdef{c#1} {\mathcal}{ABCDEFGHIJKLMNOPQRSTUVWXYZ} 
\letterdef{rm#1}{\mathrm} {dDeimM} 

\begin{document}
\begin{titlepage}
\vbox{
    \halign{#\hfil         \cr
           } 
      }  
\vspace*{1mm}
\begin{center}
{\LARGE \bf
On the D(--1)/D7-brane systems
}

\vspace*{8mm}

{\Large M.~Bill\`o$\,{}^{a}$, M.~Frau$\,{}^{a}$, F.~Fucito$\,{}^{b}$,
L.~Gallot$\,{}^{c}$, \\[1mm]
A.~Lerda$\,{}^{d}$ and J.F.~Morales$\,{}^{b}$}
\vspace*{8mm}

${}^a$ Universit\`a di Torino, Dipartimento di Fisica,\\
and I.N.F.N. - sezione di Torino,\\
		   Via P. Giuria 1, I-10125 Torino, Italy
		   \vskip 0.5cm

${}^b$ I.N.F.N. - sezione di Roma Tor Vergata\\
and Universit\`a di Roma Tor Vergata, Dipartimento di Fisica\\
	Via della Ricerca Scientifica, I-00133 Roma, Italy,\\
		   \vskip 0.5cm		
		   
${}^c$ Laboratoire d'Annecy-le-vieux de Physique Th\'eorique LAPTh\\
	Univ. Grenoble Alpes, Univ. Savoie Mont-Blanc and CNRS, \\
	F-74000, Annecy, France
	
		\vskip 0.5cm		

${}^d$  Universit\`a del Piemonte Orientale,\\
			Dipartimento di Scienze e Innovazione Tecnologica\\
			Viale T. Michel 11, I-15121 Alessandria, Italy \\
						and I.N.F.N. - sezione di Torino,\\
		   Via P. Giuria 1, I-10125 Torino, Italy
	
\vskip 0.8cm
	{\small
		E-mail:
		\texttt{billo,frau,lerda@to.infn.it; fucito,morales@roma2.infn.it;
		laurent.gallot@lapth.cnrs.fr}
	}
\vspace*{0.5cm}
\end{center}

\begin{abstract}
We study non-perturbative effects in supersymmetric U($N$) gauge theories in eight dimensions realized by means of D(--1)/D7-brane systems with non-trivial world-volume fluxes turned on.
Using an explicit string construction in terms of vertex operators, we derive the action for the
open strings ending on the D(--1)-branes and exhibit its BRST structure.
The space of vacua for these open strings is shown to be in correspondence with the moduli space
of generalized ADHM gauge connections which trigger
the non-perturbative corrections in the eight-dimensional theory. These corrections are computed
via localization and turn out to depend on the curved background used to localize the integrals on
the instanton moduli space, and vanish in flat space.
Finally, we show that for specific choices of the background the instanton partition functions reduce to weighted sums of the solid partitions of the integers.

\end{abstract}
\vskip 0.6cm
	{
		Keywords: {$\mathcal{N}=2$ SYM theories, instantons, D-branes}
	}
\end{titlepage}

\tableofcontents
\vspace{1cm}
\begingroup
\allowdisplaybreaks

\section{Introduction}
\label{secn:intro}
 
Non-perturbative effects in gauge theories can be rephrased in the language of string theory by 
considering the effects induced by branes of lower dimensions distributed along the world-volume 
of branes of higher dimensions. The prime example of this construction is represented 
by the D(--1)/D3-brane system which realizes the ADHM moduli space of gauge instantons in 
four-dimensional gauge theories 
\cite{Witten:1995im,Douglas:1996uz,Green:1997tn,Green:1998yf,Green:2000ke,Billo:2002hm, Maccaferri:2018vwo}. 
The massless sector of the open strings with at least one 
end-point on the D(--1)-branes accounts for the moduli of the gauge instanton solutions in four 
dimensions, and their effective action can be recovered by computing string scattering amplitudes on 
disk diagrams connecting the two stacks of branes \cite{Green:1997tn,Green:1998yf,Green:2000ke,Billo:2002hm}. 

This analysis can be generalized to other brane systems. In particular, new stringy non-perturbative corrections to four-dimensional gauge theories can be obtained by adding Euclidean D3-branes 
wrapping non-trivial cycles of the internal space. When orientifold planes are inserted, these exotic 
instanton configurations \cite{Blumenhagen:2006xt,Ibanez:2006da,Argurio:2006ny,Argurio:2007vqa,Bianchi:2007wy,Blumenhagen:2007zk,Ibanez:2007rs,Ibanez:2007tu} can 
generate non-perturbative effects in the effective action which are prohibited in perturbation theory, like for instance certain Majorana mass terms 
or Yukawa couplings, which may be relevant for phenomenological applications.

These methods can also be used to study non-perturbative effects in eight-dimensional field theories by considering systems of D7-branes in presence of D-instantons. Indeed, the prepotential generated by exotic instantons in O($N$) gauge theories in eight dimensions has been studied in 
\cite{Billo:2009di,Fucito:2009rs} by means of a system made of D(--1) and D7-branes 
on top of an orientifold O7-plane, allowing to explicitly test the non-perturbative 
heterotic/Type I string duality \cite{Polchinski:1995df}. 
More general D$p$/D$p^\prime$ systems can be also 
considered after turning on fluxes for the Neveu-Schwarz $B$-field, in such a way 
that supersymmetry is restored in the vacuum \cite{Witten:1995im,Witten:2000mf}.

The peculiarity of all these exotic or higher dimensional instanton systems is that they lack 
the bosonic moduli connected to the size and the gauge orientation of the instanton configurations. Moreover, they possess extra fermionic zero-modes, in addition to those connected with the broken supersymmetries, which can make their contributions to the effective action 
vanish if they are not properly lifted or removed. The problems related to the presence of these 
extra fermionic zero-modes can be cured in various ways, for example by adding orientifold or orbifold projections, or more generically by considering curved backgrounds. However,
in all these cases the non-perturbative configurations always remain point-like since 
no bosonic moduli can account for a non-zero size.
The fundamental reason for this is that in all these exotic configurations there are eight directions with mixed Dirichlet/Neumann
boundary conditions, differently from the standard cases in which there are only four mixed directions. 

The non-perturbative contributions to the gauge theory effective action are obtained after an explicit evaluation of the instanton partition functions which are expressed as integrals over the instanton moduli space. This is possible thanks to the localization
techniques introduced in \cite{Nekrasov:2002qd} for the $\mathcal{N}=2$ super 
Yang-Mills theories in four dimensions, building also on previous results in \cite{Moore:1998et}. In the end, the instanton partition functions are reduced
to integrals of rational functions which can be performed using standard complex analysis methods.
Actually, the poles of these rational functions that contribute to the integrals can be put in
one-to-one correspondence with sets of Young tableaux which in turn are related to the partitions
of the integer numbers 
\cite{Nekrasov:2002qd,Flume:2002az,Bruzzo:2002xf,Nekrasov:2003rj,Bruzzo:2003rw,Marino:2004cn}. 
When this approach is used for gauge theories in dimensions higher than four, higher dimensional Young tableaux appear and a connection with the planar and solid partitions of the integers arises \cite{Nekrasov:2003rj,Nekrasov:2017cih}. The instanton partition functions are thus expressed as weighted sums over integer, plane or solid partitions. 
Therefore, studying these systems may be interesting not only for the string theory applications we mentioned above, but also for mathematical reasons, related directly to such weighted sums (the generating function of solid partitions is not yet known in closed form) but also, for instance, in the study of Donaldson-Thomas invariants of Calabi-Yau four-folds \cite{Cao:2017swr}. 

Recently, the instanton partition functions of a pure supersymmetric U($N$) gauge theory in nine dimensions and of its conformal extension with a U($N$) flavor symmetry have been analyzed 
from this point of view in \cite{Nekrasov:2017cih,Nekrasov:2018xsb} 
and shown to be given by a simple all-instanton plethystic exponential formula. More recently, these systems have been generalized to non-conformal set-ups 
with a U($M$) flavor symmetry in \cite{Fucito:2020bjd} where also the instanton contributions to the chiral correlators have been computed. In all cases it turns out that the non-perturbative sectors 
are described by sets of moduli that share some features of both the standard and exotic instantons. Thus, in \cite{Nekrasov:2017cih,Nekrasov:2018xsb} it has been conjectured that they might arise from the open strings of a D0/D8-brane system with a Neveu-Schwarz $B$-field and with anti-D8 branes in the background.
 
The aim of this paper is to provide an explicit string theory derivation of the moduli space and of the
instanton partition functions studied in \cite{Nekrasov:2017cih,Nekrasov:2018xsb}. We consider a
D(--1)/D7-brane system in Type II B string theory 
and do not introduce any anti-branes which would make the configuration unstable, but instead turn on a magnetic flux along the world-volume of the D7-branes\,%
\footnote{Notice that this magnetic flux is an open string background and thus
is not represented by a Neveu-Schwarz $B$-field which belongs to the closed string sector.}.
More precisely, we start from a stack of $(N+M)$ D7-branes to describe a gauge theory in eight dimensions; then we introduce a constant magnetic
field on the first $N$ D7-branes to break the gauge group to $\mathrm{U}(N)\times \mathrm{U}(M)$, 
maintaining stability and supersymmetry. Finally, we add $k$ D(--1)-branes.
The open strings with at least one end-point on the D-instantons describe the
moduli space of the non-perturbative configurations. The spectrum of the open strings that start and end on the D(--1)-branes is standard and includes, among others, the bosonic moduli describing the positions of the instantonic branes in the world-volume of the seven-branes. 
The spectrum of the mixed strings stretching between the D({1) and the D7-branes is instead peculiar because, despite the presence
of eight directions with mixed Dirichlet/Neumann boundary conditions, due to the presence of the magnetic flux, it comprises a set of bosonic moduli that
can be associated to the size and orientations of the instanton configurations in eight dimensions.

We also provide a detailed analysis of the vertex operators associated to all moduli and use them to derive the effective instanton action from disk amplitudes. After introducing vacuum expectation values for the scalar fields on the D7-branes and turning on a closed-string background with Ramond-Ramond fluxes, which is known to mimic the so-called $\Omega$-background in the moduli space \cite{Billo:2006jm}, we compute the instanton partition function
using localization. Our results agree with those in \cite{Nekrasov:2017cih,Nekrasov:2018xsb};
in particular for a suitable choice of the background the instanton partition function reduces to a weighted sum over the solid partitions of the integers. Our derivation also suggests a possible consistent generalization of the results of \cite{Nekrasov:2017cih,Nekrasov:2018xsb} in which the 
$\Omega$ background is less constrained; we will investigate this possibility in a future work. 

Finally, we show that the moduli space of vacua of the matrix theory defined on the D(--1)-branes is compatible with an ADHM construction of instanton connections in eight dimensions, which has also been recently discussed in \cite{Bonelli:2020gku}. In the concluding section
we comment on the significance of our results for the eight-dimensional gauge theory, which receives
non-perturbative corrections from these instanton configurations only in its U(1) part and only in
curved space. Our notations and conventions, together with some more technical material, are
collected in the appendices.

\section{The D7-brane system and its open strings}
\label{secn:D7}
In Type II B string theory, we consider a stack of $(N+M)$ D7-branes aligned along the 
directions $\mu,\nu,\ldots=1,\ldots,8$ (with Euclidean signature). 
The directions 9 and 10 are, instead, transverse.
\begin{table}[ht]
  \begin{center}
    \begin{tabular}{cccccccc|cc} 
      1&2&3&4&5&6&7&8&9&10\\
      \hline\hline
      $-$ & $-$ & $-$ & $-$ & $-$ & $-$& $-$& $-$&$*$&$*$
      \\
    \end{tabular}
  \end{center}
  \caption{The D7-branes are aligned along the first eight directions.}
    \label{tab:table1}
\end{table}

On the world-volume of these D7-branes there is an eight-dimensional gauge theory 
with group  U($N+M)$ and sixteen supercharges. 
The part of the effective action which only depends on the gauge field strength 
can be written as 
\begin{equation}
S_{\mathrm{D7}} = S_{2}+S_{4}+\cdots~.
\label{action0}
\end{equation}
Here $S_{2}$ is the quadratic Yang-Mills action in eight dimensions
\begin{equation}
S_{2}=
\frac{1}{2 g_{\mathrm{YM}}^2}\int \!d^8x~ \mathrm{tr}\big(F^2\big)
\label{SYM}
\end{equation}
with a dimensionful gauge coupling constant
\begin{equation}
 \label{gym8}
{g_{\mathrm{YM}}^2} \equiv {4\pi g_s (2\pi\sqrt{\alpha'})^4}~,
\end{equation}
($g_s$ is the string coupling and $\sqrt{\alpha'}$ is the string length),
while $S_{4}$ is a quartic action of the form
\begin{equation}
\begin{aligned}
S_{4}=
-\frac{1}{4! \lambda^4}\int \!d^8x ~\mathrm{\tr}\big(t_8\,F^4\big) 
-\frac{\ii\,\vartheta}{4!(2\pi)^4}\int \!d^8x ~\mathrm{\tr}\big(F\wedge F\wedge 
F\wedge F\big) 
\end{aligned}
\label{action4}
\end{equation}
where
\begin{equation}
 \label{lambdadef}
{\lambda^4} \equiv {4\pi^3 g_s}
\end{equation}
is a dimensionless coupling, and $\vartheta$ is the vacuum angle, which in 
string theory is identified with the scalar field $C_0$ of the RR sector according to
\begin{equation}
\vartheta=2\pi\,C_0~.
\label{C0}
\end{equation}
The eight-index tensor $t_8$ appearing in the first term of
(\ref{action4}) is such that \cite{Tseytlin:1986ti} %
\footnote{For an explicit definition of $t_8$, see for instance Appendix B of \cite{Billo':2009gc}.}
\begin{eqnarray}
\mathrm{tr}\big(t_8 F^4\big) \! & \!\equiv \! & \! \frac{1}{16}\,t_8^{\mu_1\mu_2\cdots
\mu_7\mu_8} \,\mathrm{tr} \big(F_{\mu_1\mu_2}\cdots F_{\mu_7\mu_8}\big)
\label{f4}\\
& \!= \! &\mathrm{tr}\Big(F_{\mu\nu}F^{\nu\rho}F^{\lambda\mu}F_{\rho\lambda} 
+\frac{1}{2}\,F_{\mu\nu}F^{\rho\nu}F_{\rho\lambda}F^{\mu\lambda}
-\frac{1}{4}\,F_{\mu\nu}F^{\mu\nu}F_{\rho\lambda}F^{\rho\lambda}
-\frac{1}{8}\,F_{\mu\nu}F_{\rho\lambda}F^{\mu\nu}F^{\rho\lambda}\Big)~.
\nonumber
\end{eqnarray}
Finally, the ellipses in (\ref{action0}) stand for $\alpha^\prime$ corrections 
containing at least five field strengths or their covariant derivatives.

The gauge field strength $F_{\mu\nu}$ is actually part of a
scalar superfield
$\Phi(x,\theta)$ in eight dimensions, defined as
\begin{equation}
 \Phi(x,\theta) = \phi(x) + \sqrt{2}\,\theta\Lambda(x) +\frac{1}{2}\,
\theta\sigma^{\mu\nu}\theta\,F_{\mu\nu}(x) + \ldots
\label{Phi}
\end{equation}
where $\theta$ is the fermionic superspace coordinate, $\phi$ is a complex scalar and 
$\Lambda$ is an eight-dimensional chiral fermion\,%
\footnote{See Appendix \ref{app:notations} for our conventions on spinors and 
Dirac matrices.}.
In terms of the superfield $\Phi$, the quartic action (\ref{action4}) (plus its 
supersymmetric completion) can be written as
\begin{equation}
S_{4}= \tau\int\!d^8x\,d^8\theta~\mathrm{tr}\big(\Phi^4\big) +
\text{c.c.}
\label{quartic1}
\end{equation}
where $\tau$ is the complexified string coupling
\begin{equation}
\tau=C_0+\frac{\ii}{g_s}~.
\label{taudef}
\end{equation}
In principle the quartic action (\ref{quartic1}) can receive quantum corrections and takes the general
form
\begin{equation}
S_4^\prime=\int d^8x\,d^8\theta\,\mathcal{F}(\Phi,\tau)+\mathrm{c.c.}
\label{superpot}
\end{equation}
where $\mathcal{F}(\Phi,\tau)$ is the (holomorphic) prepotential.

We now introduce a (constant) background field on the first $N$ D7-branes.
The other $M$ D7-branes, which remain without background field, will be called from 
now on D7$^\prime$-branes. 
In this way the initial gauge symmetry group U($N+M$) is broken to U($N)$ $\times$ 
U($M$).
In particular, we consider a background that corresponds to a constant
flux on the D7-brane world-volume described by
\begin{equation}
2\pi\alpha^\prime \, F^{(0)}=\begin{pmatrix}
0& +f_1&0&0&0&0&0&0\\
-f_1&0&0&0&0&0&0&0\\
0& 0&0&+f_2&0&0&0&0\\
0& 0&-f_2&0&0&0&0&0	\\
0& 0&0&0&0&+f_3&0&0\\
0& 0&0&0&-f_3&0&0&0\\
0& 0&0&0&0&0&0&+f_4\\
0& 0&0&0&0&0&-f_4&0
\end{pmatrix}~\mathbb{1}_{N\times N}~.
\label{background}
\end{equation}
In the following we will specify the four parameters $f_1,\ldots,f_4$ in such a way to 
preserve some supersymmetry, but for the time being we may consider them as 
arbitrary\,%
\footnote{We notice that the flux breaks Lorentz invariance in the eight-dimensional 
space, so that the preserved supersymmetry can be also viewed as a two-dimensional 
supersymmetry in the transverse space.}.
The background (\ref{background}) gives rise to four different types of open strings, as
represented in Fig.~\ref{fig:1}:
the 7/7 strings starting and ending
on the $N$ D7-branes with fluxes, the 7$^\prime$/7$^\prime$ strings starting and 
ending on the $M$ D7$^\prime$-branes without fluxes, and the 7/7$^\prime$ 
or 7$^\prime$/7 strings which start and end on branes of different type.
The 7/7 and 7$^\prime$/7$^\prime$ strings are untwisted and
contain exactly the same physical states of the original strings. In particular they 
give rise to the same
effective action (\ref{action0}) for the group U($N$) (in presence of the background 
(\ref{background})) and for the group U($M$), respectively. 
The 7/7$^\prime$ or 7$^\prime$/7 strings, instead, are twisted and their
spectrum is completely different, depending on the values of the fluxes.
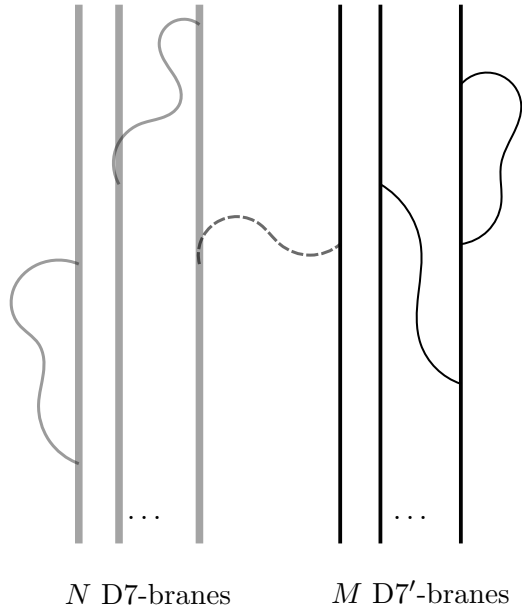
\begin{figure}
    \begin{center}
\begingroup%
  \makeatletter%
  \providecommand\color[2][]{%
    \errmessage{(Inkscape) Color is used for the text in Inkscape, but the package 'color.sty' is not loaded}%
    \renewcommand\color[2][]{}%
  }%
  \providecommand\transparent[1]{%
    \errmessage{(Inkscape) Transparency is used (non-zero) for the text in Inkscape, but the package 'transparent.sty' is not loaded}%
    \renewcommand\transparent[1]{}%
  }%
  \providecommand\rotatebox[2]{#2}%
  \newcommand*\fsize{\dimexpr\f@size pt\relax}%
  \newcommand*\lineheight[1]{\fontsize{\fsize}{#1\fsize}\selectfont}%
  \ifx\svgwidth\undefined%
    \setlength{\unitlength}{214.36789488bp}%
    \ifx\svgscale\undefined%
      \relax%
    \else%
      \setlength{\unitlength}{\unitlength * \real{\svgscale}}%
    \fi%
  \else%
    \setlength{\unitlength}{\svgwidth}%
  \fi%
  \global\let\svgwidth\undefined%
  \global\let\svgscale\undefined%
  \makeatother%
  \begin{picture}(1,1.05830163)%
    \lineheight{1}%
    \setlength\tabcolsep{0pt}%
    \put(0,0){\includegraphics{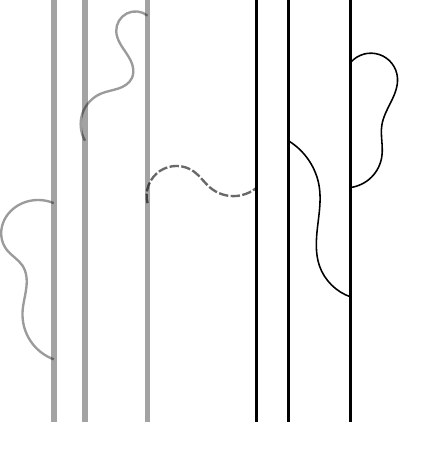}}%
    \put(0.20441477,0.15611246){\makebox(0,0)[lt]{\lineheight{1.25}\smash{\begin{tabular}[t]{l}$\ldots$\end{tabular}}}}%
    \put(0.66757272,0.15611246){\color[rgb]{0,0,0}\makebox(0,0)[lt]{\lineheight{1.25}\smash{\begin{tabular}[t]{l}$\ldots$\end{tabular}}}}%
    \put(0.09601118,0.00822731){\color[rgb]{0,0,0}\makebox(0,0)[lt]{\lineheight{1.25}\smash{\begin{tabular}[t]{l}$N$ D7-branes\end{tabular}}}}%
    \put(0.55916909,0.01005132){\color[rgb]{0,0,0}\makebox(0,0)[lt]{\lineheight{1.25}\smash{\begin{tabular}[t]{l}$M$ D7$^\prime$-branes\end{tabular}}}}%
  \end{picture}%
\endgroup%

    \end{center}
    \label{fig:d7d7p}
    \caption{
        A background flux is turned on on the $N$ D7 branes. The D7$/$D7 and 
        D7$^\prime$/D7$\prime$ open strings are untwisted, the mixed ones (depicted as 
        dashed) are 
        instead twisted.
    }
    \label{fig:1}
\end{figure} 

\subsection{The D7/D7$^\prime$ and D7$^\prime$/D7 strings}
To analyze the spectrum of the D7/D7$^\prime$ and D7$^\prime$/D7 open 
strings, it is convenient to group the space-time coordinates in pairs and introduce the 
complex combinations
\begin{equation}
z^I=\frac{x^{2I-1}+\ii \,x^{2I}}{\sqrt{2}} \quad
\mbox{and}\quad
\overbar{z}_I=\frac{x^{2I-1}-\ii \,x^{2I}}{\sqrt{2}}
\label{complexstructure}
\end{equation}
for $I=1,2,3,4,5$. The directions with $I=1,2,3,4$ are longitudinal to 
branes, while the direction with $I=5$ is transverse. In this complex notation, the 
background (\ref{background}) 
implies that in the 7/7$^\prime$ sector, the complex string coordinates $\big\{Z^I(z),
\overbar{Z}_I(z)\big\}$ and their fermionic
partners $\big\{\Psi^I(z),\overbar{\Psi}_I(z)\big\}$ with $I=1,2,3,4$ are twisted with 
a twist parameter $\theta_I$ given by\,%
\footnote{In the 7$^\prime$/7 sector we have to send $f_I$ to $-f_I$.
\label{footnotefI}}
\begin{equation}
\label{thetaf}
\rme^{2\pi\ii\,\theta_I}=\frac{1-\ii\,f_I}{1+\ii\,f_I}~.
\end{equation}
This means that they have the following monodromy properties around the origin of 
the world-sheet:
\begin{equation}
Z^I(\rme^{2\pi\ii}z)=\rme^{2\pi\ii\,\theta_I}\,Z^I(z)~,\qquad
\overbar{Z}_I(\rme^{2\pi\ii}z)=\rme^{-2\pi\ii\,\theta_I}\,\overbar{Z}_I(z)~,
\label{monodromyZ}
\end{equation}
and
\begin{equation}
\Psi^I(\rme^{2\pi\ii}z)=\pm\,\rme^{2\pi\ii\,\theta_I}\,\Psi^I(z)~,\qquad
\overbar{\Psi}_I(\rme^{2\pi\ii}z)=\pm\,\rme^{-2\pi\ii\,\theta_I}\,\overbar{\Psi}_I(z)~.
\label{monodromyPsi}
\end{equation}
Here $z$ is a point in the upper half complex plane and the sign 
in (\ref{monodromyPsi}) is 
$+$ for the Neveu-Schwarz (NS) boundary conditions and $-$ for the Ramond (R) ones.
In \cite{Bertolini:2005qh}, the bosonic and fermionic conformal 
field theory for generic twist parameters has been worked out in detail and 
we summarize the most relevant features in Appendix~\ref{app_twist}. Here we just recall
that the monodromy properties (\ref{monodromyZ}) of the bosonic coordinates 
do not allow for the existence of zero-modes associated to momentum in these directions.

The bosonic and fermionic coordinates $\big\{Z^5(z),\overbar{Z}_5(z)\big\}$ and 
$\big\{\Psi^5(z),\overbar{\Psi}_5(z)\big\}$ are instead untwisted. They have the 
standard mode expansion of untwisted fields, but with 
Dirichlet/Dirichlet boundary conditions since they are transverse to both type of branes. 
Due to these boundary conditions there are no bosonic zero-modes in these directions
either. Thus, when all four $\theta_I$'s are
different from zero, the momentum is not defined in any direction, and the 
corresponding string excitations represent non-dynamical degrees of freedom.

As mentioned above, we are interested in supersymmetric configurations. 
This means that the background (\ref{background}) must be such that
\begin{equation}
\theta_1+\theta_2+\theta_3+\theta_4 \in \mathbb{Z}~.
\label{susy}
\end{equation}
Of course, if $\theta_I=0$ for all $I$, the background vanishes and all sixteen 
supercharges are conserved on the branes. If two of the twists are zero, the system 
preserves eight supercharges while if only one
of the $\theta_I$'s vanishes the system has four conserved supercharges. 
If all twists are non-zero and satisfy (\ref{susy}), the system preserves the minimal 
amount of supersymmetry, namely only two supercharges are conserved. 
This is the case we will consider. 
For simplicity, we assume that $0<\theta_I<\frac{1}{2}$ for all $I$'s\,%
\footnote{The reason to exclude the value $\frac{1}{2}$ is merely to avoid the 
appearance of fermionic zero-modes in the NS sector.}.
With this choice, the supersymmetry condition (\ref{susy}) becomes
\begin{equation}
\theta_1+\theta_2+\theta_3+\theta_4 =1~.
\label{susy1}
\end{equation}
In this set-up, the condition that selects the physical states is
\begin{equation}
\begin{cases}
               \displaystyle{\widehat{N}=
               \frac{1}{2}\Big(1-\sum_{I=1}^4\theta_I\Big)\qquad\text{in the NS sector}~,}
               \\[3mm]
               \displaystyle{\widehat{N}=0
               \qquad\text{in the R sector}~.}
            \end{cases}
            \label{phys}
\end{equation}
where $\widehat{N}$ is the sum of the bosonic and fermionic number operators (see
(\ref{NZ}) and (\ref{Npsi})). 
Using the supersymmetry condition (\ref{susy1}), we see that both the NS
and R physical states must actually obey
\begin{equation}
\widehat{N}=0~.
\label{physical}
\end{equation}
Since the number operator is non-negative, this requirement can be satisfied only by the vacuum.

In the NS sector, the physical vacuum $|\Omega\rangle_{\mathrm{NS}}$ is related to the
$\mathrm{SL}(2,\mathbb{R})$ invariant Fock vacuum $|0\rangle$ by the action of
four bosonic twist fields $\sigma^I_{\theta_I}(z)$ of conformal dimensions
\begin{equation}
h(\sigma^I_{\theta_I})=\frac{1}{2}\,\theta_I(1-\theta_I)
\label{hsigma}
\end{equation}
and four fermionic twist fields $s^I_{\theta_I}(z)$ of conformal dimensions
\begin{equation}
h(s_{\theta_I})=\frac{1}{2}\,\theta_I^2
\label{hs}
\end{equation}
as follows \cite{Dixon:1986qv}
\begin{equation}
|\Omega\rangle_{\mathrm{NS}} = \lim_{z\to 0}\Big( \prod_{I=1}^4
\sigma^I_{\theta_I}(z)\,s^I_{\theta_I}(z)\Big)\,\rme^{-\varphi(z)}\,|0\rangle~.
\label{OmegaNS}
\end{equation}
Here $\varphi(z)$ is the bosonic field that is used to describe the superghost system in
the bosonized formalism \cite{Friedan:1985ge} and the coefficient $-1$ in the exponent
indicates that we have chosen the standard $(-1)$-superghost picture.
Using (\ref{hsigma}), (\ref{hs}) and the supersymmetry
condition (\ref{susy1}), one can check that the vertex operator creating the
physical NS vacuum out of the Fock vacuum is a conformal field of total weight 1.

The fermionic twist fields $s^I_{\theta_I}(z)$ can be written in the bosonized formalism as
\begin{equation}
s^I_{\theta_I}(z)=\rme^{\ii\,\theta_I\,\varphi^I(z)}
\label{sbosonized}
\end{equation}
where $\varphi^I(z)$ is the field that bosonizes the fermionic string coordinates
$\big\{\Psi^I(z),\overbar{\Psi}_I(z)\big\}$ according to \cite{Friedan:1985ge}
\begin{equation}
\Psi^I(z)=\rme^{\ii\,\varphi^I(z)}~,\qquad
\overbar{\Psi}_I(z)=\rme^{-\ii\,\varphi^I(z)}~.
\label{bosonization}
\end{equation}
Of course, analogous bosonization formulas hold also for the untwisted coordinates
$\big\{\Psi^5(z),\overbar{\Psi}_5(z)\big\}$. 
Using these expressions, it is easy to realize that the sum of the charges of the 
five bosonic fields $\varphi_I$ appearing in the vertex operator corresponding to
the NS vacuum minus the superghost charge is an even integer, namely
\begin{equation}
\sum_{I=1}^4\theta_I+0-(-1)=2~.
\end{equation}
Denoting by $F$ this combination of charges, we have
\begin{equation}
(-1)^F\,|\Omega\rangle_{\mathrm{NS}}=+|\Omega\rangle_{\mathrm{NS}}~.
\label{parityNS}
\end{equation}

Let us now consider the R sector. In this case the physical vacuum is created out of the
Fock vacuum by the same bosonic twist fields $\sigma_{\theta_I}(z)$ as before, but in the
fermionic part we have the following twist fields
\begin{equation}
s^I_{\theta_I-\frac{1}{2}}(z)=\rme^{\ii\,(\theta_I-\frac{1}{2})\varphi^I(z)}
\end{equation}
for $I=1,\ldots,4$ and the spin field
\begin{equation}
S^{\pm}(z)=\rme^{\pm\frac{\ii}{2}\varphi^5(z)}
\label{spinfield5}
\end{equation}
in the fifth complex untwisted direction. The presence of this spin field implies that
the R vacuum is actually a two-component SO(2) spinor. Altogether, working in 
the standard $(-\frac{1}{2})$-superghost picture, in the R sector we find two physical 
vacuum states which in the bosonized formalism are given by
\begin{equation}
|\Omega,\pm\rangle_{\mathrm{R}} = \lim_{z\to 0}\Big( \prod_{I=1}^4
\sigma^I_{\theta_I}(z)\rme^{\ii\,(\theta_I-\frac{1}{2})\varphi^I(z)}\Big)\,
S^{\pm}(z)\,\rme^{-\frac{1}{2}\varphi(z)}\,|0\rangle~.
\label{OmegaR}
\end{equation} 
The combination of charges that before we denoted by $F$ takes the value 0 on 
$|\Omega,+\rangle_{\mathrm{R}}$ and the value $-1$ on 
$|\Omega,-\rangle_{\mathrm{R}}$. Thus we have
\begin{equation}
(-1)^F\,|\Omega,\pm\rangle_{\mathrm{R}}=\pm|\Omega,\pm\rangle_{\mathrm{R}}~,
\label{parityR}
\end{equation}
showing that $(-1)^F$ actually measures the chirality of R vacuum.

If we define the GSO projector as
\begin{equation}
P_{\mathrm{GSO}}=\frac{1+(-1)^F}{2}~,
\label{PGSO}
\end{equation}
then, using (\ref{parityNS}) and (\ref{parityR}), we see that it
selects the vacuum in the NS sector and the component with positive chirality 
in the R sector, realizing in this way a supersymmetric physical spectrum containing just 
two states, the bosonic vacuum $|\Omega\rangle_{\mathrm{NS}}$ and the fermionic 
vacuum $|\Omega,+\rangle_{\mathrm{R}}$. 
We stress once more that these states do not carry any momentum and thus represent 
non-dynamical degrees of freedom since they do not propagate in space. Moreover, they 
transform in the fundamental of U($N$) and in the anti-fundamental of U($M$), since 
they arise from oriented open strings stretching between the $N$ D7-branes and the 
$M$ D7$^\prime$-branes.

We conclude this analysis with a couple of comments. First, the GSO-projection in the NS 
appears to be different with respect to the one we are used to in the standard untwisted 
case when one removes the (tachyonic) vacuum and selects states with an odd number of 
fermionic oscillators. However, if one reformulates this standard GSO projection in terms 
of $F$, one can realize that it is precisely given by (\ref{PGSO}). Secondly, the 
analysis we have described for the D7/D7$^\prime$ strings can be repeated for the 
D7$^\prime$/D7 strings without any difficulty, 
and also in that case one finds that the physical states are 
a non-dynamical scalar, corresponding to the NS vacuum, and the chiral component of a 
non-dynamical fermion, corresponding the chiral component of the R vacuum. 
Of course, these states of the 7$^\prime$/7 sector transform in the anti-fundamental 
of U($N$) and in the fundamental of U($M$).

\subsection{The no-force condition}
\label{subsecn:noforce7}
We have seen that the condition (\ref{susy1}) guarantees a supersymmetric physical 
spectrum. To complete our analysis we now show that the same condition also implies 
that the partition functions
of the mixed strings of the 7/7$^\prime$ and 7$^\prime$/7 sectors vanish, 
thus leading to a no-force condition. 

Let us first consider the 7/7$^\prime$ open strings whose partition function is
\begin{equation}
Z_{7/7^\prime}=
\int_0^\infty\!\frac{dt}{2t}~\mathrm{Tr}_{7/7^\prime}\Big(
P_{\mathrm{GSO}}\,\rme^{-2\pi t\,L_0}\Big)
\label{Z77'}
\end{equation}
where the GSO projector is defined in (\ref{PGSO}) and $L_0$ is the total open string 
Hamiltonian consisting of the orbital part and of the ghost and 
superghost parts (see for instance \cite{Friedan:1985ge} for details). 
The trace in (\ref{Z77'}) is computed over all excitations of the 7/7$^\prime$ strings 
with a $+$ sign in the bosonic NS sector and a $-$ sign in the fermionic R sector. This 
calculation is pretty standard (see for example \cite{Billo:1998vr,Billo':2009gc} and 
references therein), but we would like to point out just a couple of subtle issues that 
arise in the present case. First of all, in the 7/7$^\prime$ sector, there are eight (real) 
twisted directions and two (real) untwisted directions. This is precisely the case in which 
the odd spin-structure $\mathrm{R}(-1)^F$ gives 
a non-vanishing contribution to the string partition function due to a cancellation 
between the fermionic zero-modes in the two untwisted directions and the bosonic 
zero-modes of the superghost-sector 
\cite{Bergman:1997rf,Bergman:1997gf,Billo:1998vr}. Secondly, the presence
of the background (\ref{background}) along the D7-brane world-volume
modifies the contribution of the zero-modes of the eight twisted bosonic coordinates by 
multiplying the eight-dimensional volume $V_8$ by the factor $\prod_I f_I$.

Taking this into account and using the GSO projection (\ref{PGSO}), we find
\begin{equation}
Z_{7/7^\prime}=\frac{NM\,V_8}{(4\pi^2\alpha^\prime)^4}\,\prod_{I=1}^4f_I\!
\int_{0}^\infty\!\frac{dt}{2t}~\frac{1}{2}
\left[\prod_{I=1}^4\frac{\vartheta_3(\theta_It)}{\vartheta_1(\theta_It)}+
\prod_{I=1}^4\frac{\vartheta_4(\theta_It)}{\vartheta_1(\theta_It)}-
\prod_{I=1}^4\frac{\vartheta_2(\theta_It)}{\vartheta_1(\theta_It)}-
1\right]
\label{Z77'a}
\end{equation}
where the factor $NM$ accounts for the color multiplicity of each string excitation and
the four terms in the square brackets correspond to the contributions of the
NS, NS$(-1)^F$, R and R$(-1)^F$ spin-structures, respectively. Here, as usual, we have 
defined:
\begin{equation}
\vartheta_1(z)\equiv\vartheta\sp{1}{1}\big(\ii z|\ii t)~,~~
\vartheta_2(z)\equiv\vartheta\sp{1}{0}\big(\ii z|\ii t)~,~~
\vartheta_3(z)\equiv\vartheta\sp{0}{0}\big(\ii z|\ii t)~,~~
\vartheta_4(z)\equiv\vartheta\sp{0}{1}\big(\ii z|\ii t)
\label{thetas}
\end{equation}
where $\vartheta\sp{a}{b}\big(\ii z|\ii t)$ are the Jacobi $\vartheta$-functions with
characteristics (see for example \cite{Green:1987mn}). They satisfy the Jacobi-Riemann 
identity
\begin{equation}
\prod_{I=1}^4 \vartheta_3(z_I)-\prod_{I=1}^4 \vartheta_4(z_I)-
\prod_{I=1}^4 \vartheta_2(z_I)-\prod_{I=1}^4 \vartheta_1(z_I)=
-2\prod_{I=1}^4\vartheta_1(Z/2- z_I)
\end{equation}
with $Z=\sum_I z_I$. Applying this identity in our case where $z_I=\theta_It$ and 
using the periodicity properties of the Jacobi $\vartheta$-functions, we can 
recast the previous identity in the following form
\begin{equation}
\prod_{I=1}^4 \vartheta_3(\theta_It)+\prod_{I=1}^4 \vartheta_4(\theta_It)-
\prod_{I=1}^4 \vartheta_2(\theta_It)-\prod_{I=1}^4 \vartheta_1(\theta_It)=0~,
\label{identity}
\end{equation}
when the twist parameters satisfy the supersymmetry relation (\ref{susy1}).
Thus, inserting this relation into (\ref{Z77'a}), we find
\begin{equation}
Z_{7/7^\prime}=0~.
\label{Z77'b}
\end{equation}
In a completely similar way one can show that also for the other orientation the partition
function vanishes:
\begin{equation}
Z_{7^\prime/7}=0~.
\label{Z7'7}
\end{equation}
This fact, together with the vanishing of the partition function in the 7/7 and 
$7^\prime/7^\prime$ sectors due to the BPS property of these branes, 
implies that our brane system is stable when the
condition (\ref{susy1}) is satisfied, and that one can pile up an arbitrary number 
of D7 and D7$^\prime$-branes since the net force in all channels is zero.

\section{Adding D--instantons}
\label{secn:dinstantons}
To study some non-perturbative features in the effective theory defined on the 
world-volume of the D7 and D7$^\prime$-branes,
we add a stack of $k$ D--instantons, {\it{i.e.}} D-branes
with Dirichlet boundary conditions in all directions also known as D$(-1)$-branes.
This addition gives rise to new open string sectors: the (--1)/(--1) strings starting
and ending on the D-instantons, the (--1)/7 and 7/(--1) strings which stretch
between the D-instantons and the D7-branes, and finally 
(--1)/7$^\prime$ and 7$^\prime$/(--1) strings which connect the D-instantons
and the D7$^\prime$-branes. 
We now analyze the physical states in all these sectors.

\subsection{The D(--1)/D(--1) strings}

The open strings stretching between two D(--1)-branes are untwisted and have
Dirichlet/Dirichlet boundary conditions in all ten directions. Thus, they do not carry
any momentum and describe non-propagating degrees of freedom. 
To describe the physical states in this sector, it is convenient 
to distinguish the eight directions, that are longitudinal to the D7 and D7$^\prime$-
branes from the two ones that are transverse, and organize the open string states 
in representations of~Spin(8)$\,\times\,$Spin(2).

In the NS sector, the only physical states are those with one fermionic oscillator, corresponding to 
the ten-dimensional vector. 
Using the same complex notation introduced in the previous section,
we can pack the eight (real) components of this vector along the D7 or 
D7$^\prime$-branes into four complex variables which we denote by $B^I$, 
together with their complex conjugates $\overbar{B}_I$, with $I=1,2,3,4$. 
Their corresponding vertex operators, in the $(-1)$-superghost picture, are\,%
\footnote{The normalization factors in these and in all the following vertices are chosen for later convenience so that the final formulas look simpler.}
\begin{equation}
V_{B^I}=\frac{B^I}{\sqrt{2}}\,\overbar{\Psi}_I(z)\,\rme^{-\varphi(z)}
\quad\mbox{and}\quad 
V_{\overbar{B}_I}=\frac{\overbar{B}_I}{\sqrt{2}}\,\Psi^I(z)\,\rme^{-\varphi(z)}~.
\label{VBI}
\end{equation}
The remaining two components of the vector, along the transverse directions to the 
branes, give rise to the variables $\xi$ and $\chi$, 
whose vertex operators are
\begin{equation}
V_{\xi}=\xi\,\overbar{\Psi}_5(z)\,\rme^{-\varphi(z)}
\quad\mbox{and}\quad 
V_{\chi}=\frac{\chi}{2}\,\Psi^5(z)\,\rme^{-\varphi(z)}~.
\label{Vchi}
\end{equation}
All these vertex operators are conformal fields of dimension 1 and possess an even 
$F$-charge, as one can see using the bosonization formulas (\ref{bosonization}); thus they 
are preserved by the GSO projection (\ref{PGSO}).

In the R sector the only physical state is the vacuum. This state is actually a sixteen-component spinor 
of Spin(10), which the GSO projection fixes to be anti-chiral. To describe these sixteen
components we use the formalism of spin fields \cite{Friedan:1985ge} and introduce the
notation
\begin{equation}
S^{\pm\pm\pm\pm}(z)=\rme^{\pm\frac{\ii}{2}\varphi^1(z)}\,
\rme^{\pm\frac{\ii}{2}\varphi^2(z)}
\rme^{\pm\frac{\ii}{2}\varphi^3(z)}
\rme^{\pm\frac{\ii}{2}\varphi^4(z)}
\label{spinfields}
\end{equation} 
to denote the spin fields in the first four complex directions, which have 
to be used together with the spin fields $S^\pm(z)$ in the fifth direction
already introduced in (\ref{spinfield5}).
The physical ground state of the R sector, in the $(-\frac{1}{2})$-superghost picture, 
is therefore given by
\begin{equation}
\lim_{z\to 0} S^{\pm\pm\pm\pm}(z)\,S^{\pm}(z)\,\rme^{-\frac{1}{2}\varphi(z)}\,
|0\rangle
\end{equation}
with a total odd number of $-$ signs in the spin fields
in order to have an even $F$-charge and be preserved
by the GSO projection. 
It is convenient to distinguish the cases in which the last spin field is 
$S^+(z)$ or $S^-(z)$. 

If we have $S^+(z)$, the spin fields in the first four directions 
must have an odd number of minus signs. Clearly, there are eight possibilities. 
Using a notation that resembles the one used for the components of the vector field in 
the NS sector, we denote these eight spin fields by $S^I(z)$ and $\overbar{S}_I(z)$
with $I=1,\ldots,4$ (see (\ref{SI}) for their detailed definition).
The components of the fermionic R ground state corresponding to these fields
are denoted by $\overbar{M}_I$ and $M^I$, respectively, and are described by the following
vertex operators of weight 1:
\begin{equation}
V_{M^I}=M^I\,\overbar{S}_I(z)\,S^+(z)\,\rme^{-\frac{1}{2}\varphi(z)}
\quad\mbox{and}\quad 
V_{\overbar{M}_I}=S^I(z)\,S^+(z)\,\rme^{-\frac{1}{2}\varphi(z)}\,\overbar{M}_I~.
\label{VMI}
\end{equation}

If instead we have $S^-(z)$, the spin fields in the first four directions must have an even 
number of minus signs and there are again eight possibilities. We denote these spin fields by
$S(z)$, $\overbar{S}(z)$ and $S^{IJ}(z)=-S^{JI}(z)$ with $I,J=1,\ldots,4$
(see (\ref{SIJ}) for details). We can also introduce the conjugate fields $\overbar{S}_{IJ}(z)$, but 
they are not independent since the following relation holds
\begin{equation}
\overbar{S}_{IJ}(z)=\frac{1}{2}\,\epsilon_{IJKL}\,S^{KL}(z)~.
\label{Sbar}
\end{equation}
Using this notation, we see that the eight components of the R ground 
states with $S^-(z)$, which we denote by $\lambda^{IJ}$, $\lambda$ and $\eta$,
are described by the following vertex operators
\begin{equation}
V_{\lambda^{IJ}}=\frac{\lambda^{IJ}}{\sqrt{2}}\,\overbar{S}_{IJ}(z)\,S^-(z)\,
\rme^{-\frac{1}{2}\varphi(z)}~,
\label{VlambdaIJ}
\end{equation}
and
\begin{equation}
V_{\lambda}=\lambda\,\frac{\overbar{S}(z)-S(z)}{\sqrt{2}\,\ii}\,S^-(z)\,\rme^{-\frac{1}{2}\varphi(z)}~,
\quad
V_{\eta}=\eta\,\frac{\overbar{S}(z)+S(z)}{\sqrt{2}}\,S^-(z)\,\rme^{-\frac{1}{2}\varphi(z)}~.
\label{Vlambda}
\end{equation}
It is easy to verify that these vertex operators are conformal fields of weight 1. 
Furthermore, the fields $\lambda$ and $\eta$ are real, 
while the fields $\lambda^{IJ}$ are complex 
with their complex conjugates 
$\overbar{\lambda}_{IJ}$ given by
\begin{equation}
\overbar{\lambda}_{IJ}=\frac{1}{2}\,\epsilon_{IJKL}\,\lambda^{KL}~,
\label{lambdabaris}
\end{equation}
as a consequence of the relation (\ref{Sbar}).

Since there are $k$ D-instantons, all variables appearing as polarizations
in the above vertex operators are $(k\times k)$ matrices transforming in the adjoint 
representation of U($k$).

\subsection{The D(--1)/D7 and D7/D(--1) strings}
We now consider the open strings stretching between the 
D-instantons and the D7-branes. In this case there are two facts that one has to 
consider, namely that the first four complex space directions have mixed 
Dirichlet/Neumann boundary conditions and that there is a background field on 
the D7-branes given by (\ref{background}).
The combined net effect is that the first four complex directions have an 
effective twist parameter given by $\big(\frac{1}{2}-\theta_I\big)$.

Taking this into account, one can see that the condition which 
selects the physical states in the NS sector is
\begin{equation}
\widehat{N}+\frac{1}{2}-\frac{1}{2}\,\sum_{I=1}^4\theta_I=0
\end{equation}
where $\widehat{N}$ is the total number operator. Inserting the constraint
(\ref{susy1}), we see that the physical states in the NS sector must have $\widehat{N}=0$, which
is possible only for the vacuum. Thus in the spectrum we have only one state whose
polarization we denote by $w$. Its corresponding vertex operator, in the 
$(-1)$-superghost picture, is
\begin{equation}
V_w=\frac{w}{\sqrt{2}}\,\Big(\prod_{I=1}^4\sigma^I_{\frac{1}{2}-\theta_I}(z)~
\rme^{\ii(\frac{1}{2}-\theta_I)\varphi^I(z)}\Big)\,\rme^{-\varphi(z)}
\label{Vw}
\end{equation}
which is a conformal field of weight 1 when the condition (\ref{susy1}) is satisfied.
The same condition also guarantees that the total $F$-charge is even so that this
vertex is preserved by the GSO projection.

In the R sector, the condition for physical states implies $\widehat{N}=0$. Thus, also 
in this sector the vacuum is the only physical state. Due to the R boundary conditions
in the untwisted fifth complex direction, the ground state is degenerate and is a 
two-component spinor denoted by $\mu^{\pm}$. 
The corresponding vertex operator in the $(-\frac{1}{2})$-superghost picture is
\begin{equation}
V_{\mu^\pm}=\mu^\pm\,\Big(\prod_{I=1}^4\sigma^I_{\frac{1}{2}-\theta_I}(z)~
\rme^{-\ii\theta_I\varphi^I(z)}\Big)\,S^{\pm}(z)\,\rme^{-\frac{1}{2}\varphi(z)}
\label{Vmu}
\end{equation}
which is a conformal field of weight 1.
Exploiting the supersymmetry condition (\ref{susy1}), it is easy to see that
\begin{equation}
(-1)^F\,V_{\mu^\pm}=\pm\,V_{\mu^\pm}
\end{equation}
so that only $\mu^+$ is selected by the GSO projection. To simplify the notation, we
will denote $\mu^+$ simply as $\mu$.

This analysis can be easily repeated for the strings with the opposite orientation
stretching between the D7-branes and the D-instantons. The result is that in the NS sector 
we have only one GSO-even physical state $\overbar{w}$ whose 
vertex operator is
\begin{equation}
V_{\overbar{w}}=\frac{\overbar{w}}{\sqrt{2}}\,\Big(\prod_{I=1}^4\sigma^I_{\theta_I-\frac{1}{2}}(z)~
\rme^{\ii(\theta_I-\frac{1}{2})\varphi^I(z)}\Big)\,\rme^{-\varphi(z)}~,
\label{Vbarw}
\end{equation}
while in the R sector the only GSO-even physical state is $\overbar{\mu}$ whose vertex
operator is
\begin{equation}
V_{\overbar{\mu}}=\overbar{\mu}\,\Big(\prod_{I=1}^4\sigma^I_{\theta_I-\frac{1}{2}}(z)~
\rme^{\ii\theta_I\varphi^I(z)}\Big)\,S^{+}(z)\,\rme^{-\frac{1}{2}\varphi(z)}~.
\label{Vmubar}
\end{equation}
Taking into account the multiplicity of the D-instantons and of the D7-branes, the 
variables $w$ and $\mu$ are $(k\times N)$ matrices transforming in the 
($\mathbf{k},\overbar{\mathbf{N}}$) representation 
of $\mathrm{U}(k)\times\mathrm{U}(N)$, while
the conjugate variables $\overbar{w}$ and $\overbar{\mu}$ are ($N\times k)$ 
matrices transforming in the ($\overbar{\mathbf{k}},\mathbf{N}$) representation of the
same group.

\subsection{The D(--1)/D7$^\prime$ and D7$^\prime$/D(--1) strings}

We now consider the strings connecting the D-instantons and the D7$^\prime$-branes.
The only but important difference with the previous case is that now there is no 
background field on the 7-branes so that the string fields in the first four complex 
directions have standard Dirichlet/Neumann boundary conditions without extra twists.
Therefore, we can formally set $\theta_I=0$ in the previous formulas. If we do this, we
immediately realize that in the NS sector the physical states must satisfy the condition
\begin{equation}
\widehat{N}+\frac{1}{2}=0~,
\end{equation}
which has no solution. Thus, there are no physical states in the NS part of the spectrum.
This fact was already observed in \cite{Billo':2009gc} where the D$(-1)$/D7-brane system
was explored in connection with the first studies of exotic instantons in eight 
dimensions.

In the R sector, instead, the condition for physical states is satisfied by
the degenerate ground state that describes a two-component spinor which
we denote by $\mu^{\prime\,\pm}$. The corresponding vertex operator is
\begin{equation}
V_{\mu^{\prime\pm}}=\mu^{\prime\pm}
\,\Big(\prod_{I=1}^4\sigma^{I}_{\frac{1}{2}}(z)\Big)\,S^{\pm}(z)\,\rme^{-\frac{1}{2}\varphi(z)}
\label{Vmuprime}
\end{equation}
where $\sigma^I_{\frac{1}{2}}$ is the appropriate bosonic twist field when the
$I$-th coordinate has Dirichlet/Neumann boundary conditions. This twist field
has conformal dimension $\frac{1}{8}$, and thus the vertex operator 
(\ref{Vmuprime}) has total weight 1. Furthermore, it is easy to check 
\begin{equation}
(-1)^F\,V_{\mu^{\prime\pm}}=\mp\,V_{\mu^{\prime\pm}}~,
\end{equation}
so that only $\mu^{\prime-}$, which we will simply call $\mu^\prime$, 
survives the GSO projection.

For the strings with opposite orientation that stretch between the D7$^\prime$-branes
and the D-instantons, we find a similar structure with no physical states in the NS sector, 
and only one physical state in the R sector which we denote $\overbar{\mu}^\prime$.
Its vertex operator is
\begin{equation}
V_{\overbar{\mu}^{\prime}}=\overbar{\mu}^{\prime}
\,\Big(\prod_{I=1}^4{\sigma}^{I}_{-\frac{1}{2}}(z)\Big)\,S^{-}(z)
\,\rme^{-\frac{1}{2}\varphi(z)}
\label{Vmubarprime}
\end{equation}
where ${\sigma}^{I}_{-\frac{1}{2}}$ is the bosonic twist field corresponding to an $I$-th
coordinate with Neumann/Dirichlet boundary conditions, that is conjugate to the
twist field $\sigma^I_{\frac{1}{2}}$.

Taking into account the multiplicity of the branes, we have that $\mu^\prime$ is
a $(k\times M)$ matrix transforming the $(\mathbf{k},\mathbf{\overbar{M}})$ 
representation of $\mathrm{U}(k)\times\mathrm{U}(M)$, while 
$\overbar{\mu}^\prime$ is $(M\times k)$ matrix transforming the 
$(\mathbf{\overbar{k}},\mathbf{M})$ representation of this group.

\subsection{D-instanton partition functions}
We now study the partition function of the open strings with at least one end-point on the D-instantons. In the $(-1)/(-1)$ sector, we simply have
\begin{equation}
Z_{(-1)/(-1)}=0
\label{Z-1-1}
\end{equation}
as a consequence of the BPS condition satisfied by the D(--1)-branes.

Let us now consider the $(-1)/7^\prime$ and the $7^\prime/(-1)$ sectors. Recalling that 
the D7$^\prime$-branes do not have any background field on their world-volume, the calculation
of the partition functions $Z_{(-1)/7^\prime}$ and $Z_{7^\prime/(-1)}$ is exactly the same as the
one described in detail in \cite{Billo':2009gc}, which we briefly recall here. For the strings starting
from the D-instantons we have
\begin{equation}
Z_{(-1)/7^\prime}=kM
\int_{0}^\infty\!\frac{dt}{2t}~\frac{1}{2}
\left[\Big(\frac{\vartheta_2(0)}{\vartheta_4(0)}\Big)^4
+\Big(\frac{\vartheta_1(0)}{\vartheta_4(0)}\Big)^4-\Big(\frac{\vartheta_3(0)}{\vartheta_4(0)}\Big)^4
-1\right]
\label{Z-17'a}
\end{equation}
where the prefactor accounts for the multiplicity of the branes, and the four terms inside the square
brackets arise from the four spin structures with eight directions with Dirichlet/Neumann 
boundary conditions. We notice in particular that the second term, corresponding to the NS$(-1)^F$
sector, actually vanishes since $\vartheta_1(0)=0$, while the fourth term, corresponding to the
R$(-1)^F$ spin structure, is non-zero due to the cancellation of the fermionic zero-modes in the
two directions with Dirichlet/Dirichlet boundary conditions with the bosonic zero-modes of the
superghost system \cite{Billo:1998vr}. Exploiting the abstruse identity
\begin{equation}
\vartheta_3(0)^4-\vartheta_4(0)^4-\vartheta_2(0)^4=0~,
\label{abstruse}
\end{equation}
we easily conclude that
\begin{equation}
Z_{(-1)/7^\prime}=-kM\int_{0}^\infty\!\frac{dt}{2t}~.
\label{Z-17'b}
\end{equation}
The partition function for the other orientation corresponds to placing the D-instantons 
``on the other
side'' of the 7-brane, which is obtained with a parity transformation that reverses the sign of the
odd spin structure R$(-1)^F$ \cite{Billo:1998vr}. Then we have
\begin{equation}
Z_{7^\prime/(-1)}=kM
\int_{0}^\infty\!\frac{dt}{2t}~\frac{1}{2}
\left[\Big(\frac{\vartheta_2(0)}{\vartheta_4(0)}\Big)^4
+\Big(\frac{\vartheta_1(0)}{\vartheta_4(0)}\Big)^4-\Big(\frac{\vartheta_3(0)}{\vartheta_4(0)}\Big)^4
+1\right]~.
\label{Z7'-1a}
\end{equation}
The abstruse identity (\ref{abstruse}) now implies that\,%
\footnote{In the T-dual set-up, namely in the D0/D8-brane system, the vanishing of the partition function can be interpreted as due to the creation of a fundamental string, see for example
\cite{Bergman:1997gf,Kitao:1998vn,Billo:1998vr}.}
\begin{equation}
Z_{7^\prime/(-1)}=0~.
\label{Z7'-1b}
\end{equation}

Let us now consider the contribution of the open strings connecting the D-instantons to the 
D7-branes. In this case, due to the presence of the background (\ref{background}) on the 7-branes,
we have an extra twist to take into account with respect to the previous cases. In particular,
the partition function $Z_{(-1)/7}$ is obtained from (\ref{Z-17'a}) by twisting the 
$\vartheta$-functions by $-\theta_It$ in each complex direction and replacing $M$ by $N$, which 
is the number of D7-branes. In other words we have
\begin{equation}
Z_{(-1)/7}=kN
\int_{0}^\infty\!\frac{dt}{2t}~\frac{1}{2}
\left[\prod_{I=1}^4\frac{\vartheta_2(-\theta_It)}{\vartheta_4(-\theta_It)}+
\prod_{I=1}^4\frac{\vartheta_1(-\theta_It)}{\vartheta_4(-\theta_It)}-
\prod_{I=1}^4\frac{\vartheta_3(-\theta_It)}{\vartheta_4(-\theta_It)}-
1\right]~.
\label{Z-17a}
\end{equation}
Since the twists satisfy the condition (\ref{susy1}), the Jacobi-Riemann identity (\ref{identity}), 
together with the following parity properties
\begin{equation}
\vartheta_1(-z)=-\vartheta_1(z)~,\quad\vartheta_2(-z)=+\vartheta_2(z)~,\quad
\vartheta_3(-z)=+\vartheta_3(z)~,\quad\vartheta_4(-z)=+\vartheta_4(z)~,
\end{equation}
implies that
\begin{equation}
Z_{(-1)/7}=0~.
\label{Z-17b}
\end{equation}
The partition function for the other orientation is obtained from (\ref{Z7'-1a})
by twisting the $\vartheta$-functions with $\theta_It$ and replacing $M$ with $N$, namely
\begin{equation}
Z_{7/(-1)}=kN
\int_{0}^\infty\!\frac{dt}{2t}~\frac{1}{2}
\left[\prod_{I=1}^4\frac{\vartheta_2(\theta_It)}{\vartheta_4(\theta_It)}+
\prod_{I=1}^4\frac{\vartheta_1(\theta_It)}{\vartheta_4(\theta_It)}-
\prod_{I=1}^4\frac{\vartheta_3(\theta_It)}{\vartheta_4(\theta_It)}+
1\right]~.
\label{Z7-1a}
\end{equation}
Using the Jacobi-Riemann identity, we easily find
\begin{equation}
Z_{7/(-1)}=kN
\int_{0}^\infty\!\frac{dt}{2t}~.
\label{Z7-1b}
\end{equation}

Summing all contributions (\ref{Z-1-1}), (\ref{Z-17'b}), (\ref{Z7'-1b}), (\ref{Z-17b}) and
(\ref{Z7-1b}), we conclude that the total partition function of the open strings with at least one
end-point on the D-instantons is
\begin{equation}
\begin{aligned}
Z_{\mathrm{tot}}=k(N-M)\int_{0}^\infty\!\frac{dt}{2t}~.
\end{aligned}
\end{equation}
If we set 
\begin{equation}
N=M~,
\label{N=M}
\end{equation}
the total partition function vanishes and no forces are present.
This no-force condition is equivalent to requiring that the brane system be conformal invariant %
\footnote{If $N\neq M$ the total open-string partition function with D-instantons 
does not vanish and the gauge theory on the D7-brane becomes not-conformal in the sense that the dimensionless coupling constant $\lambda$ of the quartic action 
(\ref{action4}) runs with a $\beta$-function coefficient proportional to $(N-M)$. Furthermore, as we will show in later sections, when $N\neq M$ the instanton partition function $Z_k$
is not dimensionless and the instanton counting parameter $q$ carries dimensions of mass$^{(N-M)}$, in such a way that $q^k\,Z_k$ is dimensionless.}.

\section{The instanton moduli and their action}
\label{secn:moduliaction}
In the previous sections we have described a system made of $N$ D7-branes with flux, 
$M$ D7$^\prime$-branes without flux and $k$ D-instantons. The flux 
on the D7-branes is subject to the condition (\ref{susy1}) with all $\theta_I$ different 
from zero in order to preserve the minimal amount of supersymmetry. 
For simplicity, from now on we choose the twists to be all equal, namely
\begin{equation}
\theta_1=\theta_2=\theta_3=\theta_4=\frac{1}{4}~.
\label{theta14}
\end{equation}
With this choice the four complex directions are treated symmetrically and many
formulas simplify. For example, the fermionic twist fields to be used in the vertex operators of
the (--1)/7 strings (see (\ref{Vw}) and (\ref{Vmu})) become
\begin{equation}
\Sigma(z)=\rme^{\frac{\ii}{4}(\varphi^1(z)+\varphi^2(z)+\varphi^3(z)+\varphi^4(z))}
\quad\mbox{and}\quad
\overbar{\Sigma}(z)=\rme^{-\frac{\ii}{4}(\varphi^1(z)+\varphi^2(z)+\varphi^3(z)+\varphi^4(z))}
\label{SigmabarSigma}
\end{equation}
in the NS and R sectors, respectively. The same twist fields are used also in the vertex
operators of the 7/(--1) strings (see (\ref{Vbarw}) and (\ref{Vmubar})), but in this case $\Sigma(z)$
is used in the R sector and $\overbar{\Sigma}(z)$ in the NS sector. Both $\Sigma(z)$ and 
$\overbar{\Sigma}(z)$ are conformal fields of weight $\frac{1}{8}$. It is also convenient to
define the bosonic twist fields
\begin{equation}
\Delta(z)=\prod_{I=4}\sigma^I_{\frac{1}{4}}(z)
\quad\mbox{and}\quad
\overbar{\Delta}(z)=\prod_{I=4}\sigma^I_{-\frac{1}{4}}(z)
\label{DeltabarDelta}
\end{equation}
to be used for the (--1)/7 and 7/(--1) strings respectively. These fields have conformal 
dimension $\frac{3}{8}$ and are conjugate to each other. Similarly, for the strings of type 
(--1)/7$^\prime$ and 7$^\prime$/(--1) we define the following pair of conjugate twist fields
\begin{equation}
\Delta^\prime(z)=\prod_{I=4}\sigma^I_{\frac{1}{2}}(z)
\quad\mbox{and}\quad
\overbar{\Delta}^\prime(z)=\prod_{I=4}\sigma^I_{-\frac{1}{2}}(z)~,
\end{equation}
which have conformal weight $\frac{1}{2}$.

Using these notations, we summarize for future reference
all physical states of the open strings with at least one end-point on the D-instantons and 
their vertex operators in Table~\ref{table:moduli}.

\begin{table}[ht]
\begin{center}
\begin{tabular}{ |c|c|c|c|c| }
\hline
Sector & Moduli & $\phantom{\Big|}$Vertex Operators &Statistics& Dimensions\\ \hline\hline
\multirow{8}{*}{$(-1)/(-1)$} & $B^I~,~\overbar{B}_I$ 
& $\phantom{\bigg|}\frac{B^I}{\sqrt{2}}\,\overbar{\Psi}_I\,\rme^{-\varphi}~,~
\phantom{\bigg|}\frac{\overbar{B}_I}{\sqrt{2}}\,\Psi^I\,\rme^{-\varphi}$& bosonic&(length)$^{-1}$\\
& $\xi~,~\chi$ 
& $\phantom{\bigg|}\xi\,\overbar{\Psi}_5\,\rme^{-\varphi}~,~
\phantom{\bigg|}\frac{\chi}{2}\,\Psi^5\,\rme^{-\varphi}$& bosonic&(length)$^{-1}$\\
&$M^I~,~\overbar{M}_I$ 
& $\phantom{\bigg|}M^I\,\overbar{S}_I\,S^+\,\rme^{-\frac{1}{2}\varphi}~,~
\phantom{\bigg|}S^I\,S^+\,\rme^{-\frac{1}{2}\varphi}\,\overbar{M}_I$
& fermionic&(length)$^{-\frac{3}{2}}$\\
 & $\lambda^{IJ}$ 
& $\phantom{\bigg|}\frac{\lambda^{IJ}}{\sqrt{2}}\,\overbar{S}_{IJ}\,S^-\,\rme^{-\frac{1}{2}\varphi}$& fermionic
&(length)$^{-\frac{3}{2}}$\\
& $\lambda$ 
& $\phantom{\bigg|}\lambda\,\frac{\overbar{S}-S}{\sqrt{2}\,\ii}\,S^-\,\rme^{-\frac{1}{2}\varphi}$& fermionic&(length)$^{-\frac{3}{2}}$\\
& $\eta$ 
& $\phantom{\bigg|}\eta\,\frac{\overbar{S}+S}{\sqrt{2}}\,S^-\,\rme^{-\frac{1}{2}\varphi}$& fermionic&(length)$^{-\frac{3}{2}}$\\
 \hline
\multirow{3}{*}{$(-1)/7$} & $w$ & $\phantom{\bigg|}\frac{w}{\sqrt{2}}\,
\Delta \,\Sigma\,\rme^{-\varphi}$ 
& bosonic&(length)$^{-1}$\\
& $\mu$ & $\phantom{\bigg|}\mu\,\Delta \,\overbar{\Sigma}\,S^+\,\rme^{-\frac{1}{2}\varphi}$ 
& fermionic&(length)$^{-\frac{3}{2}}$\\
\hline
\multirow{3}{*}{$7/(-1)$} & $\overbar{w}$ & $\phantom{\bigg|}\frac{\overbar{w}}{\sqrt{2}}\,
\overbar{\Delta} \,\overbar{\Sigma}\,\rme^{-\varphi}$ & bosonic&(length)$^{-1}$\\
& $\overbar{\mu}$ & $\phantom{\bigg|}\overbar{\mu}\,
\overbar{\Delta} \,\Sigma\,S^+\,\rme^{-\frac{1}{2}\varphi}$ 
& fermionic&(length)$^{-\frac{3}{2}}$\\
\hline
$(-1)/7^\prime$ & $\mu^\prime$ & $\phantom{\bigg|}
\mu^\prime\,
\Delta^\prime\,S^-\,\rme^{-\frac{1}{2}\varphi}$ & fermionic&(length)$^{-\frac{3}{2}}$\\
\hline
$7^\prime/(-1)$ & $\overbar{\mu}^\prime$ & $\phantom{\bigg|}
\overbar{\mu}^\prime\,
\overbar{\Delta}^\prime\,S^-\,\rme^{-\frac{1}{2}\varphi}$ & fermionic&(length)$^{-\frac{3}{2}}$\\
\hline
\end{tabular}
\end{center}
\caption{The physical moduli corresponding to open strings with at least one end-point on 
the D-instantons, and their vertex operators in the canonical superghost pictures, $(-1)$ in the NS
sector and $(-\frac{1}{2})$ in the R sector. The last two columns contain their statistics and their
scaling dimensions.}
\label{table:moduli}
\end{table}

\subsection{The moduli action}
The action of the $(-1)/(-1)$ moduli can be obtained by dimensionally reducing 
the $\mathcal{N}=1$ super Yang-Mills action from ten to zero dimensions and, in the
notations previously introduced, it reads as follows
\begin{align}
S_{(-1)}&=\frac{1}{g_0^2} \,\mathrm{tr}\,\bigg\{\,\frac{1}{2}
 \big[B^I,\overbar{B}_I\big]\big[B^J,\overbar{B}_J\big]-
 \big[B^I,B^J\big]\big[\overbar{B}_I,\overbar{B}_J\big]
 -\big[\xi,B^I\big] \big[\chi,\overbar{B}_I\big]
-\big[\chi,B^I\big]\big[\xi,\overbar{B}_I\big]
\notag\\[1mm]
&\quad
+\frac{1}{2}\big[\chi,\xi\big]^2+\frac{1}{2}\,\lambda \big[\chi,\lambda\big]
+\frac{1}{2}\,\eta \big[\chi,\eta\big]
+\frac{1}{4}\,\lambda^{IJ}\big[\chi,\lambda^{KL}\big]\,\epsilon_{IJKL}\notag
\\[1mm]
&\quad-2\,M^I\big[\xi,\overbar{M}_I\big]
+\lambda^{IJ}\Big(\big[\overbar{B}_I,\overbar{M}_J\big]-
\big[\overbar{B}_J,\overbar{M}_I\big]+\epsilon_{IJKL}\,\big[B^K,M^L\big]\Big)
\notag\\[1mm]
&\quad+(\eta+\ii\,\lambda)\big[\overbar{B}_I,M^I\big]
+(\eta-\ii\,\lambda)\big[B^I,\overbar{M}_I\big]
\bigg\}\label{S-1-1}
\end{align}
where 
\begin{equation}
g_0^2=\frac{g_s}{4\pi^3\alpha^{\prime\,2}}
\end{equation}
is the Yang-Mills coupling constant in zero dimensions.

The action involving the moduli of the mixed sectors is
\begin{equation}
S_{\mathrm{mixed}}=\frac{1}{g_0^2} \,\mathrm{tr}\,\big\{
\overbar{w}\,(\chi\,\xi+\xi\,\chi)\,w -2\,\overbar{\mu}\,\xi\,\mu
-\overbar{\mu}\,(\eta+\ii\,\lambda)\, w
-\overbar{w}\,(\eta-\ii\,\lambda)\,\mu+\overbar{\mu}^\prime\,
\chi\,\mu^\prime\big\}~.
\label{S-17}
\end{equation}
Using standard conformal field theory methods, one can verify that all cubic couplings in the 
above actions can be obtained by computing the 3-point amplitudes among the vertex operators
introduced in the previous section. If one applies this method to compute the 4-point 
amplitudes and obtain the quartic couplings, one encounters divergent integrals over the vertex
insertion points, as it has also been recently pointed out in \cite{Sen:2020cef}. To circumvent this 
problem, we adopt the strategy already used in \cite{Billo:2002hm,Billo:2006jm,Billo':2009gc} 
and introduce auxiliary fields to disentangle the quartic interactions. For our purposes it is enough 
to consider the first two quartic terms in (\ref{S-1-1}). The minimal set of auxiliary fields
which are needed to decouple these interactions comprises seven auxiliary fields, denoted 
by $D$ and $D^{IJ}$ with $D^{JI}=-D^{IJ}$ ($I,J=1,\ldots,4$). 
Notice that $D$ is real while the six $D^{IJ}$'s are complex with their complex conjugates
$\overbar{D}_{IJ}$ given by
\begin{equation}
\overbar{D}_{IJ}=\frac{1}{2}\,\epsilon_{IJKL}\,D^{KL}~.
\label{Dbaris}
\end{equation}
Thus $D^{IJ}$ correspond to six real degrees of freedom. Using these auxiliary 
fields, to which we assign canonical scaling dimensions of (length)$^{-2}$, 
the first two terms of (\ref{S-1-1}) can be 
replaced by 
\begin{equation}
\begin{aligned}
\frac{1}{g_0^2}\,\mathrm{tr}\,\bigg\{&\!\frac{1}{4}\,D^{IJ}D^{KL}\,\epsilon_{IJKL}-
D^{IJ}\Big(\big[\overbar{B}_I,\overbar{B}_J\big]+\frac{1}{2}\,\epsilon_{IJKL}\,\big[B^K,B^L\big]\Big)
+\frac{1}{2}\,D^2-\ii\, D\big[B^I,\overbar{B}_I\big]
\bigg\}~.
\end{aligned}
\label{SDD}
\end{equation}
Indeed, eliminating the auxiliary fields through their equations of motion and using the Jacobi identity, one can show that (\ref{SDD}) reduces exactly to the first two terms of (\ref{S-1-1}).

We observe that the cubic terms in (\ref{SDD}) can be obtained by computing the couplings among the vertex operators $V_{B^I}$ and $V_{\overbar{B}_I}$, given in (\ref{VBI}),
and the following vertex operators for the auxiliary fields
\begin{equation}
\begin{aligned}
V_{D^{IJ}}&=-\frac{D^{IJ}}{2}\Big(\overbar{\Psi}_I(z)\overbar{\Psi}_J(z)+\frac{1}{2}\,
\epsilon_{IJKL}\,\Psi^K(z)\Psi^L(z)\Big)~,\\
V_D&=-\frac{\ii\,D}{2}\,\Psi^I(z)\overbar{\Psi}_I(z)~.
\end{aligned}
\end{equation}
These vertices are in the $(0)$-superghost picture, as appropriate for auxiliary fields, and are
conformal operators of weight 1.

The vertex $V_D$ has a non-vanishing coupling also with the vertex operators $V_{w}$
and $V_{\overbar{w}}$ of the mixed bosonic moduli $w$ and $\overbar{w}$; this 
coupling leads to the following term 
\begin{equation}
\frac{1}{g_0^2}\,\mathrm{tr}\,\big\{\!\!-\ii\,D\,w\,\overbar{w}\big\}
\label{Dwwbar}
\end{equation}
that has to be added to the moduli action (\ref{S-17}). Furthermore,
we find convenient to add also the term
\begin{equation}
\frac{1}{g_0^2} \,\mathrm{tr}\,\big\{\,\overbar{h}^\prime\,h^\prime\,\big\}
\label{S-17pp}
\end{equation}
where $h^\prime$ and its conjugate $\overbar{h}^\prime$ are auxiliary fields with scaling
dimensions of (length)$^{-2}$. 
Even if these fields  look trivial since they do not interact with any other moduli, it is useful
to introduce them for reasons that will be clear in a moment. Also these auxiliary 
fields can be described by vertex operators in the $(0)$-superghost picture, which are 
\begin{equation}
V_{h^\prime}= h^\prime\,\Delta^\prime(z)\,\frac{\overbar{S}(z)+S(z)}{\sqrt{2}}
\quad\mbox{and}\quad
V_{\overbar{h}^\prime}= \overbar{h}^\prime\,
\overbar{\Delta}^\prime(z)\,\frac{\overbar{S}(z)+S(z)}{\sqrt{2}}~.
\end{equation}
These operators are conformal fields of weight 1.

We now introduce ADHM-like variables by means of the following rescalings
\begin{equation}
B^I\to g_0\,B^I~,~~
M^I\to g_0\,M^I~,~~
w\to g_0\,w~,~~
\mu\to g_0\,\mu~,~~
\mu^\prime\to g_0\,\mu^\prime~,~~
h^\prime\to g_0\,h^\prime
\label{rescalings}
\end{equation}
with analogous ones for their conjugates, in such a way that the rescaled bosons $B^I$ and $w$ have dimensions of length, the rescaled fermions $\mu$ and $\mu^\prime$ have dimensions of
(length)$^{\frac{1}{2}}$, and the rescaled auxiliary field $h^\prime$ becomes dimensionless.
After these rescalings, adding all contributions 
one finds that the instanton moduli action can be written as
\begin{equation}
S_{\text{inst}}=\frac{1}{g_0^2}\,S_G+S_K+S_D
\label{Sinstfinal}
\end{equation}
where
\begin{subequations}
\begin{align}
S_G&=\mathrm{tr}\,\bigg\{\,
\frac{1}{4}\,D^{IJ} D^{KL}\,\epsilon_{IJKL}+\frac{1}{2}\,D^2+\frac{1}{2}\big[\chi,\xi\big]^2
+\frac{1}{2}\,\lambda \big[\chi,\lambda\big]
\notag\\
&\quad\qquad
+\frac{1}{2}\,\eta \big[\chi,\eta\big]
+\frac{1}{4}\,\lambda^{IJ}\big[\chi,\lambda^{KL}\big]\,\epsilon_{IJKL}\bigg\}
~,\label{SG}\\[1mm]
S_K&=\mathrm{tr}\,\bigg\{\!\!-\big[\xi,B^I\big] \big[\chi,\overbar{B}_I\big]
-\big[\chi,B^I\big]\big[\xi,\overbar{B}_I\big]
-2\,M^I\big[\xi,\overbar{M}_I\big]\notag\\
&\quad\qquad-2\,\overbar{\mu}\,\xi\,
\mu+\overbar{w}\,\xi\,\chi\,w+\overbar{w}\,\chi\,\xi\,w+\overbar{\mu}^\prime\,\chi
\,\mu^\prime+\overbar{h}^\prime\,
h^\prime\bigg\}~,
\label{SK}\\[1mm]
S_D&=\mathrm{tr}\,\bigg\{\!\!-\ii\, D\Big(\big[B^I,\overbar{B}_I\big]+w\,\overbar{w}\Big)
-D^{IJ}\Big(\big[\overbar{B}_I,\overbar{B}_J\big]+\frac{1}{2}\,\epsilon_{IJKL}
\,\big[B^K,B^L\big]\Big)\notag\\
&\qquad\quad
+\lambda^{IJ}\Big(\big[\overbar{B}_I,\overbar{M}_J\big]-\big[\overbar{B}_J,\overbar{M}_I\big]
+\epsilon_{IJKL}\,\big[B^K,M^L\big]\Big)\notag\\
&\qquad\quad
+(\eta+\ii\,\lambda)\Big(\big[\overbar{B}_I,M^I\big]+w\,\overbar{\mu}\Big)
+(\eta-\ii\,\lambda)\Big(\big[B^I,\overbar{M}_I\big]-\mu\,\overbar{w}\Big)
\bigg\}~.
\label{SD}
\end{align}
\label{SGSKSD}%
\end{subequations}
In the field theory limit $\alpha^\prime\to 0$, or equivalently in the strong-coupling limit $g_0\to\infty$, 
the term $S_G$ can be discarded and the fields $D$ and $D^{IJ}$
become Lagrange multipliers for the bosonic constraints
\begin{equation}
\big[B^I,\overbar{B}_I\big]+w\,\overbar{w}=0~,\qquad
\big[\overbar{B}_I,\overbar{B}_J\big]+\frac{1}{2}\,\epsilon_{IJKL}\,\big[B^K,B^L\big]=0~,
\label{bosconstraints}
\end{equation} 
while the fields $(\eta\pm\ii\,\lambda)$ and $\lambda^{IJ}$ become Lagrange multipliers for
the fermionic constraints
\begin{align}
\big[\overbar{B}_I,M^I\big]+w\,\overbar{\mu}=0~,~~~
\big[\overbar{B}_I,\overbar{M}_J\big]-\big[\overbar{B}_J,\overbar{M}_I\big]
+\epsilon_{IJKL}\,\big[B^K,M^L\big]=0~.
\label{fermconstraints}
\end{align}
In Section~\ref{secn:ADHM} we will show that the same equations arise from the
ADHM construction for instantons in eight dimensions. This fact allows us to regard
the open strings ending on the D(--1) branes as instanton moduli.

\subsection{Introducing vacuum expectation values and the $\varepsilon$-background}
We now generalize the action (\ref{Sinstfinal}) by introducing the
interactions with external backgrounds.

From the string theory point of view, these backgrounds correspond either 
to vacuum expectation values of massless fields propagating on the world-volume of the D7 and 
D7$^\prime$-branes, or to vacuum expectation values of bulk fields in the closed 
string sectors. Therefore, one can obtain their interactions with the instanton 
moduli by following the procedure described in 
\cite{Billo:2002hm,Billo:2006jm,Billo:2009di}, namely by computing the
correlation functions among the vertex operators of the moduli and the vertex operators 
of the external background fields.
However, there is also an alternative route to obtain these interactions which exploits the symmetries of the brane system.

Let us give some details. The moduli actions (\ref{SGSKSD}) are invariant by construction 
under the symmetry group U($k$) of the D-instantons, and the moduli $\chi$ and $\xi$ can be interpreted as the parameters of infinitesimal U($k$) transformations\,%
\footnote{The use of $\chi$ or $\xi$ is fixed by a neutrality condition with respect to the charge in the fifth complex plane of the initial string construction.}. In our conventions, these 
transformations are
\begin{equation}
\mathsf{T}_{\chi}[\,\bullet\,]=\begin{cases}
-[\chi, \bullet]~~\mbox{if $\bullet$ is an adjoint modulus}~,\\
-\chi \bullet~~\mbox{if $\bullet$ is a fundamental modulus}~,\\
~\bullet \chi~~\mbox{if $\bullet$ is an anti-fundamental modulus}~,
\end{cases}
\end{equation}
and similarly for $\mathsf{T}_{\xi}$.
Using this notation, for example the terms $-\big[\xi,B^I\big] \big[\chi,\overbar{B}_I\big]$ 
and $\overbar{\mu}^\prime\,\chi\,\mu^\prime$ appearing in (\ref{SK}) can be rewritten 
respectively as
\begin{equation}
-\mathsf{T}_{\xi}[\,B^I\,]\,\mathsf{T}_{\chi}[\,\overbar{B}_I\,]\quad~\mbox{and}\quad
-\overbar{\mu}^\prime\,\mathsf{T}_{\chi}[\,\mu^\prime\,]~.
\label{terms}
\end{equation}
All other $\chi$- and $\xi$-dependent terms in (\ref{SGSKSD}) can be treated in the same way.

The idea is to extend this approach to all symmetries of the brane system. 
The actions (\ref{SGSKSD}) are invariant under the U($N$) transformations
of the $N$ D7-branes and under the U($M$) transformations of the $M$ D7$^\prime$-branes, in which only the mixed moduli transform according to the representations defined 
by the open string construction. The actions (\ref{SGSKSD}) are also invariant under the
SU(4) transformations related to the rotations in the four complex planes
indexed by $I$. Let us denote by $\varepsilon_I$ the parameters of such rotations
and define the quantities
\begin{equation}
\varepsilon=\varepsilon_1+\varepsilon_2+\varepsilon_3+\varepsilon_4~,
\label{varepsilon}
\end{equation}
and
\begin{equation}
\varepsilon_{IJ}=\varepsilon_I+\varepsilon_J-\frac{\varepsilon}{2}~.
\label{epsilonIJ}
\end{equation}
Then, it is easy to show that the actions (\ref{SGSKSD}) are invariant under the 
following transformations
\begin{equation}
\begin{aligned}
(B^I,M^I)\quad&\to\quad\rme^{+\ii\,\varepsilon_I}\,(B^I,M^I)~,\\[1mm]
(\overbar{B}_I,\overbar{M}_I)\quad&\to\quad\rme^{-\ii\,\varepsilon_I}\,
(\overbar{B}_I,\overbar{M}_I)~,\\[1mm]
(\lambda^{IJ},D^{IJ})\quad&\to\quad\rme^{+\ii\,\varepsilon_{IJ}}\,
(\lambda^{IJ},D^{IJ})~,
\end{aligned}
\label{SU(4)}
\end{equation}
with all other moduli unchanged, provided $\varepsilon=0$. 
This is a SU(4) symmetry whose origin is quite clear. 
Indeed,
the initial SO(8) symmetry of the D-brane system is broken to SO(7) on the D-instantons
\cite{Moore:1998et,Billo:2009di}, and this symmetry is further reduced to 
SO(6)$\,\simeq\,$SU(4) by the complex structure we have introduced. 
Therefore, the four parameters $\varepsilon_I$ subject to the condition
$\varepsilon=0$ can be understood as the parameters of the SU(4) transformations
along the Cartan directions (see Appendix~\ref{app:notations}).

We now exploit the U($N$), U($M$) and SU(4) symmetries and replace the U($k$) transformations
$\mathsf{T}_{\chi}$ and $\mathsf{T}_{\xi}$ according to 
\begin{equation}
\begin{aligned}
\mathsf{T}_{\chi}[\,\bullet\,]~&\to~
\mathsf{T}_{(\chi,a,m,\varepsilon_I)}[\,\bullet\,]
~,\\[1mm]
\mathsf{T}_{\xi}[\,\bullet\,]~&\to~
\mathsf{T}_{(\xi,\overline{a},\overbar{m},\overline{\varepsilon}_I)}[\,\bullet\,]
\end{aligned}
\end{equation}
where $\mathsf{T}_{(\chi,a,m,\varepsilon_I)}$ and $\mathsf{T}_{(\xi,\overline{a},\overbar{m},\overline{\varepsilon}_I)}$
denote $\mathrm{U}(k)\times\mathrm{U}(N)\times
\mathrm{U}(M)\times\mathrm{SU}(4)$ infinitesimal transformations parametrized 
by $\chi$, $a$, $m$ and
$\varepsilon_I$, and by $\xi$, $\overline{a}$, $\overbar{m}$ and $\overline{\varepsilon}_I$
respectively, in the appropriate representations.
For example, applying this rule the terms in (\ref{terms}) get shifted and become
\begin{equation}
\begin{aligned}
-\mathsf{T}_{(\xi,\overline{a},\overbar{m},\overline{\varepsilon}_I)}[\,B^I\,]\,
\mathsf{T}_{(\chi,a,m,\varepsilon_I)}[\,\overbar{B}_I\,]
&=-\big(\!-\big[\xi,B^I\big] +\overline{\varepsilon}_I\,B^I\big)\big(\!-\big[\chi,\overbar{B}_I\big]-\varepsilon_I\overbar{B}_I\big)~,\\[2mm]
-\overbar{\mu}^\prime\,\mathsf{T}_{(\chi,a,m,\varepsilon_I)}[\,\mu^\prime\,]
&=-\overbar{\mu}^\prime
\,(-\chi\,\mu^\prime+\mu^\prime\,m)~.
\end{aligned}
\label{shifts}
\end{equation}
Proceeding systematically in this way, we can generate new terms in the moduli action. 
Taking for simplicity the U($N$) and U($M$) parameters only along the Cartan directions, 
these new terms 
correspond to shifting the actions (\ref{SGSKSD}) as follows\,%
\footnote{Notice that no dependence on $\overbar{m}$ arises with this 
method, since there are no terms depending both on $\xi$ and on the primed moduli 
in the original action.}
\begin{subequations}
\begin{align}
S_G~&\to~S^\prime_G=S_G+\tr\,\bigg\{\sum_{I<J , K<L}\!\!
\varepsilon_{IJ}\,\lambda^{IJ}\,\lambda^{KL}\,
\epsilon_{IJKL}\bigg\}~,\label{newSG}\\
S_K~&\to~S^\prime_K=S_K+\mathrm{tr}\,\bigg\{2 \sum_{I} \Big(
\varepsilon_{I}\,B^I\,[\xi,\overbar{B}_I]+
\overline{\varepsilon}_{I}\,B^I\,[\chi,\overbar{B}_I]+\overline{\varepsilon}_I
\,\varepsilon_I\,B^I\,\overbar{B}_I-\overline{\varepsilon}_I\,M^I\,\overbar{M}_I
\Big)\notag\\
&\qquad\qquad\qquad
-2\,a\,\overbar{w}\,\xi\,w-2\,\overbar{a}\,\overbar{w}\,(\chi-a)\,w+2\,\overbar{a}\,
\overbar{\mu}\,\mu-\overbar{\mu}^\prime\,\mu^\prime\,m\bigg\}~,\label{newSK}\\
S_D~&\to~S^\prime_D=S_D~.\label{newSD}
\end{align}
\label{newSGSKSD}%
\end{subequations}
As mentioned above, these new terms can also be obtained from the interactions
that the instanton moduli have with the background fields on the 
D7 and D7$^\prime$-branes or with the bulk fields of the closed string sectors.
In particular, the terms with $a$, $\overbar{a}$ and $m$ arise when a vacuum
expectation value is given to the adjoint scalars $\phi$ and $\overbar{\phi}$ on the
$N$ D7-branes (see (\ref{Phi})) and to the analogue scalars on the $M$ D$7^\prime$-branes. 
One can explicitly check that all such terms arise from the correlators among the vertex operators
of the instanton moduli and the vertex operators of these adjoint scalars. The latter have the
same structure as the vertices of $\xi$ and $\chi$ in (\ref{Vchi}), and thus it is not surprising that
they produce similar couplings in the moduli action. 

The terms in (\ref{newSGSKSD}) that depend on $\varepsilon_I$ and $\overline{\varepsilon}_I$
can instead be generated by introducing a graviphoton background corresponding to the
3-form field strength $\mathcal{W}=d(C_{2}-\tau\,B_{2})$ of Type II B string theory\,%
\footnote{Here $C_{2}$ and $B_{2}$ are the 2-forms in the R-R and NS-NS sectors, respectively,
and $\tau$ is the axio-dilaton (\ref{taudef}).}. In our set-up, the 3-form field strength has both 
holomorphic and anti-holomorphic components along the fifth complex direction, namely it is of the form $\mathcal{W}_{\mu\nu z}$ and $\overbar{\mathcal{W}}_{\mu\nu\overbar{z}}$ and is skew-symmetric along the $\mu$ and $\nu$ directions\,%
\footnote{As we will show later in Section~\ref{secn:BRST}, 
if $\sum_{I}\varepsilon_I=0$ the BRST structure of the moduli action is deformed but not spoiled by the background, meaning that supersymmetry is not broken.
Notice that in complex notation, the 3-form $\mathcal{W}$ in (\ref{W}) is of type (2,1), while the 
3-form $\overbar{\mathcal{W}}$ in (\ref{Wbar}) of type (1,2).}
:
\begin{subequations}
\begin{align}
&\mathcal{W}_{12 z}=\varepsilon_1~,\quad
\mathcal{W}_{34 z}=\varepsilon_2~,\quad
\mathcal{W}_{56 z}=\varepsilon_3~,\quad
\mathcal{W}_{78 z}=\varepsilon_4~,\label{W}\\
&\overbar{\mathcal{W}}_{12 \overbar{z}}=\overline{\varepsilon}_1~,\quad
\overbar{\mathcal{W}}_{34 \overbar{z}}=\overline{\varepsilon}_2~,\quad
\overbar{\mathcal{W}}_{56 \overbar{z}}=\overline{\varepsilon}_3~,\quad
\overbar{\mathcal{W}}_{78 \overbar{z}}=\overline{\varepsilon}_4~.\label{Wbar}
\end{align}
\label{WbarW}%
\end{subequations}
As shown in \cite{Billo:2006jm}, to obtain the $\varepsilon_I$-dependent terms of the moduli action, it is enough to consider the R-R part of the 3-form field-strength $\mathcal{W}$, whose vertex
operator was discussed in detail in \cite{Billo:2009di}. In the notation for the spin fields we are adopting now, the vertex operator for the holomorphic components is proportional to
\begin{equation}
\begin{aligned}
V_{\mathcal{W}}(z,\overbar{z})&=\sum_{I} \Big(\varepsilon_{I}-\frac{\varepsilon}{2}\Big)\,
\Big[\overbar{S}_I(z)\,S^+(z)\,\rme^{-\frac{1}{2}\varphi(z)}\times
\widetilde{S}^I(\overbar{z})\,\widetilde{S}^+
(\overbar{z})\,\rme^{-\frac{1}{2}\widetilde{\varphi}(\overbar{z})}\\
&\qquad\qquad\quad-
S^I(z)\,S^+(z)\,\rme^{-\frac{1}{2}\varphi(z)}\times\widetilde{\overbar{S}}_I(\overbar{z})\,
\widetilde{S}^+
(\overbar{z})\,\rme^{-\frac{1}{2}\widetilde{\varphi}(\overbar{z})}\Big]
\end{aligned}
\label{vertexRR}
\end{equation}
where the tilded and untilded operators denote the right and left-moving parts
respectively. On the other hand, the vertex for the anti-holomorphic part is proportional to
\begin{equation}
\begin{aligned}
V_{\overline{\mathcal{W}}}(z,\overbar{z})&=\sum_{I<J}
\overline{\varepsilon}_{IJ}\,
\Big[\overbar{S}_{IJ}(z)\,S^-(z)\,\rme^{-\frac{1}{2}\varphi(z)}\times
\widetilde{S}^{IJ}(\overbar{z})\,\widetilde{S}^-
(\overbar{z})\,\rme^{-\frac{1}{2}\widetilde{\varphi}(\overbar{z})}\Big]
\end{aligned}
\label{vertexRR1}
\end{equation}
where $\overline{\varepsilon}_{IJ}$ is defined in analogy with (\ref{epsilonIJ}). 
Setting $\varepsilon=\overline{\varepsilon}=0$ and computing
mixed open/closed string amplitudes on disks with insertions of the moduli vertex operators
on the boundary and of the vertex operators $V_{\mathcal{W}}$ and
$V_{\overline{\mathcal{W}}}$ in the interior\,%
\footnote{See for example \cite{Billo:2004zq} for details on mixed amplitudes involving R-R
backgrounds.}, one can correctly reproduce all terms in (\ref{newSGSKSD}) that depend on
$\varepsilon_I$ and $\overline{\varepsilon}_I$. If we release the constraint $\varepsilon=0$, the
following extra term
\begin{equation}
\frac{1}{g_0^2}\,\mathrm{tr}\,\big\{\ii\,\varepsilon\,(\lambda\,\eta-D\,\xi)\big\}
\label{newterm3}
\end{equation}
is generated by the vertex (\ref{vertexRR}).
As we will see in the next section, this term plays a crucial role when one integrates over the instanton moduli space and its presence turns out to be necessary in the intermediate steps, even if the end result will not exhibit any {\emph{explicit}} dependence on
$\varepsilon$. The full moduli action we will consider is therefore
\begin{equation}
S_{\text{inst}}^\prime=\frac{1}{g_0^2}\,\Big(S^\prime_G+\mathrm{tr}\,\big\{\ii\,\varepsilon\,(\lambda\,\eta-D\,\xi)\big\}\Big)+S^\prime_K+S^\prime_D
\label{Sprimeinstfinal}
\end{equation}
with $S^\prime_G$, $S^\prime_K$ and $S^\prime_D$ given in (\ref{newSGSKSD}).

\subsection{The BRST structure of the moduli spectrum}
\label{secn:BRST}
We now show that all moduli, except for $\chi$, can be organized in doublets with
respect to a BRST charge $\mathcal{Q}$ which is a real combination of two super-symmetries preserved in the brane system. More specifically, we take
\begin{equation}
\mathcal{Q}=\frac{Q+\overbar{Q}}{\sqrt{2}}
\label{BRST}
\end{equation}
where
\begin{equation}
Q=\oint \frac{dz}{2\pi\ii}\,S(z)\,S^-(z)\,\rme^{-\frac{1}{2}\varphi(z)}
\qquad\mbox{and}\qquad
\overbar{Q}=\oint \frac{dz}{2\pi\ii}\,\overbar{S}(z)\,S^-(z)\,\rme^{-\frac{1}{2}\varphi(z)}
\label{QbarQ}
\end{equation}
with $S(z)$ and $\overbar{S}(z)$ given in (\ref{SbarS}). 

To obtain a BRST transformation of the form
\begin{equation}
\delta X\,\equiv\, \mathcal{Q}\,X=Y~,
\end{equation}
we can follow the methods described in \cite{Billo:2002hm,Billo:2009di}, namely
we evaluate the commutator of $\mathcal{Q}$ with the vertex operator for $Y$ and from the resulting expression read the form of the BRST variation according to
\begin{equation}
\Big[\mathcal{Q},V_{Y}(w)\Big]=V_{\delta X}(w)~.
\label{QV}
\end{equation}
We can explicitly verify that there is no fermionic vertex operator which under the action of $\mathcal{Q}$ can reproduce the structure of the vertex for $\chi$. This implies that
\begin{equation}
\mathcal{Q}\,\chi=0~,
\label{BRSTchi}
\end{equation}
so that the interpretation of $\chi$ as a parameter for the U($k$) transformations can be maintained.
On the contrary, we find
\begin{equation}
\mathcal{Q}\,\xi=\eta~.
\end{equation}
The choice of the BRST charge $\mathcal{Q}$ introduces therefore an asymmetry between $\chi$ and $\xi$, which up to this point played a quite similar role.

To obtain the BRST transformation of $Y$, we impose the nilpotency of $\mathcal{Q}$ up to symmetries, namely
\begin{equation}
\mathcal{Q}\,Y=\mathcal{Q}^2\,X=\mathsf{T}_{(\chi,a,m,\varepsilon_I)}[\,X\,]~.
\end{equation}
Following this method, for the moduli of the $(-1)/(-1)$ sector we explicitly find
\begin{equation}
\begin{aligned}
&\mathcal{Q}\,B^I=M^I~,\qquad
\mathcal{Q}\,M^I=-\big[\chi,B^I\big]+\varepsilon_I\,B^I~,\\
&\mathcal{Q}\,\overbar{B}_I=\overbar{M}_I~,\qquad
\mathcal{Q}\,\overbar{M}_I=-\big[\chi,\overbar{B}_I\big]-\varepsilon_I\,\overbar{B}^I~,\\
&\mathcal{Q}\,\lambda^{IJ}=D^{IJ}~,\quad~
\mathcal{Q}\,D^{IJ}=-\big[\chi,\lambda^{IJ}\big]+\varepsilon_{IJ}\,\lambda^{IJ}~,\\
&\mathcal{Q}\,\lambda=D~,\qquad\quad~
\mathcal{Q}\,D=-\big[\chi,\lambda\big]~,\\
&\mathcal{Q}\,\xi=\eta~,\qquad\quad~~
\mathcal{Q}\,\eta=-\big[\chi,\xi\big]~.
\end{aligned}
\label{Q-1-1}
\end{equation}
with the understanding the repeated indices are not summed. For the mixed moduli, instead, we
obtain
\begin{equation}
\begin{aligned}
&\mathcal{Q}\,w=\mu~,\qquad
\mathcal{Q}\,\mu=-\chi\,w+w\,a~,\\
&\mathcal{Q}\,\overbar{w}=\overbar{\mu}~,\qquad
\mathcal{Q}\,\overbar{\mu}=\overbar{w}\,\chi-a\,\overbar{w}~,\\
&\mathcal{Q}\,\mu^{\prime}=h^\prime~,\quad~~
\mathcal{Q}\,h^{\prime}=-\chi\,\mu^{\prime}+\mu^{\prime}\,m~,\\
&\mathcal{Q}\,\overbar{\mu}^{\prime}=\overbar{h}^\prime~,\quad~~
\mathcal{Q}\,\overbar{h}^{\prime}=\overbar{\mu}^{\prime}\,\chi-m\,\overbar{\mu}^{\prime}~.
\end{aligned}
\label{Qmixed}
\end{equation}
These BRST transformations allow us to build doublets of the form $(X,\mathcal{Q}\,X)$. They 
are
\begin{equation}
(B^I,M^I)~,~~~(\overbar{B}_I,\overbar{M}_I)~,~~~
(\xi,\eta)~,~~~(\lambda^{IJ},D^{IJ})~,~~~(\lambda,D)
\label{adjointdoublets}
\end{equation}
in the $(-1)/(-1)$ sector, and
\begin{equation}
(w,\mu)~,~~~(\overbar{w},\overbar{\mu})~,~~~(\mu^\prime,h^\prime)~,~~~
(\overbar{\mu}^\prime,\overbar{h}^\prime)
\label{mixeddoublets}
\end{equation}
in the mixed sectors. 
Furthermore, we can check that the action (\ref{Sprimeinstfinal}) is BRST invariant
and $\mathcal{Q}$-exact. Indeed,
\begin{equation}
\mathcal{Q}\,S^\prime_{\mathrm{inst}}=0~,\quad
S^\prime_{\mathrm{inst}} = \mathcal{Q}\,\Lambda
\label{QS}
\end{equation}
where
\begin{equation}
\begin{aligned}
\Lambda &= \mathrm{tr}\,\bigg\{\,\frac{1}{g_0^2}\,\Big(
\frac{1}{4}\,\lambda^{IJ} D^{KL}\,\epsilon_{IJKL}+\frac{1}{2}\,\lambda\,D-\frac{1}{2}\,\eta 
\big[\chi,\xi\big]-\ii\,\varepsilon\,\lambda\,\xi\Big)\\
&\qquad~~~+\xi \Big(\big[B^I,\overbar{M}_I\big]+\big[\overbar{B}_I,M^I\big]+
w\,\overbar{\mu}-\mu\,\overbar{w}\Big)
+\overline{a}\big(\overbar{w}\,\mu-\overbar{\mu}\,w\big)
\\
&\qquad~~~-\lambda^{IJ}\Big(\big[\overbar{B}_I,\overbar{B}_J\big]+\frac{1}{2}\,\epsilon_{IJKL}\,\big[B^K,B^L\big]\Big)-\ii\,\lambda\Big(\big[B^I,\overbar{B}_I\big]+w\,\overbar{w}\Big)\\
&\qquad~~~
-\sum_I \overline{\varepsilon}_I\,\big(B^I\,\overbar{M}_I-M^I\,\overbar{B}_I\big)
+\frac{1}{2}\,\mu^\prime\,\overbar{h}^\prime+\frac{1}{2}\,
h^\prime\,\overbar{\mu}^\prime\bigg\}~.
\end{aligned}
\label{Lambdais}
\end{equation}
We observe that the dependence of the moduli action 
on $\overline{a}$, $\overline{\varepsilon}_I$ and $\varepsilon$ is entirely encoded in the fermion
$\Lambda$ on which the BRST charge acts. This fact will be important in the following, and in
particular it suggests that the end-result of the localization process will not carry any
{\emph{explicit}} dependence on these parameters.

\section{The instanton partition function}
\label{secn:partitionfunction}
The effects of the D-instantons in the D7/D7$^\prime$ system are encoded in the gran-canonical
partition function
\begin{equation}
Z=\sum_{k=0}^\infty q^k\,Z_k
\label{Zgrancanonical}
\end{equation}
where $q=\rme^{2\pi\ii\tau}$ is the instanton fugacity and $Z_k$ is the partition
function in the $k$-instanton sector (with the convention that $Z_0=1$).
The latter is defined by the following integral
\begin{equation}
Z_k=\int\!d\mathcal{M}_k~\rme^{-S^\prime_{\mathrm{inst}}}~.
\label{Zk}
\end{equation}
where $\mathcal{M}_{k}$ denotes the set of moduli with $k$ instantons. 
Actually, as we have seen in the previous section, all such moduli except for $\chi$ are paired in BRST doublets each one containing a boson and a fermion. We have in fact the following structure
\begin{equation}
d\mathcal{M}_k=d\chi(dB^I\,dM^I)(d\overbar{B}_I \,d\overbar{M}_I)
(d\lambda^{IJ}\,dD^{IJ})
(d\lambda\,dD)(d\xi\,d\eta)(dw\,d\mu)(d\overbar{w}\,d\overbar{\mu})
(d\mu^\prime\, dh^\prime)(d\overbar{\mu}^\prime\, d\overbar{h}^\prime)~.
\end{equation}
Therefore, any rescaling of the doublets cancels between bosons and fermions and
does not affect the end result. We exploit this feature and rescale all doublets 
with a factor of $1/x$.
In the limit $x\to\infty$ a drastic simplification occurs since only the quadratic terms of the action 
$S^\prime_{\mathrm{inst}}$ are relevant in the limit, 
while the cubic and quartic terms become subleading and can be neglected. In other words the moduli action becomes
\begin{align}
S^\prime_{\mathrm{inst}}&=\frac{1}{x^2g_0^2}\,\mathrm{tr}\,\bigg\{\,
\frac{1}{4}\,D^{IJ} D^{KL}\,\epsilon_{IJKL}+\frac{1}{2}\,D^2+\frac{1}{2}\big[\chi,\xi\big]^2
+\frac{1}{2}\,\lambda \big[\chi,\lambda\big]+\frac{1}{2}\,\eta \big[\chi,\eta\big]
\notag\\
&\qquad\qquad\quad
+\frac{1}{4}\,\lambda^{IJ}\big[\chi,\lambda^{KL}\big]\,\epsilon_{IJKL}+
\sum_{I<J , K<L}\!\!
\varepsilon_{IJ}\,\lambda^{IJ}\,\lambda^{KL}\,
\epsilon_{IJKL}+\ii\,\varepsilon\,(\lambda\,\eta-D\,\xi)
\bigg\}\notag\\
&~~~+\frac{1}{x^2}\,\mathrm{tr}\,\bigg\{\overbar{\mu}^\prime\,\big(\chi-m\big)
\,\mu^\prime+\overbar{h}^\prime\,
h^\prime-2\,\overbar{a}\,\big(\overbar{w}\,(\chi-a)\,w-\overbar{\mu}\,\mu\big)
\label{Sinst2}\\
&\qquad\qquad\quad
+2 \sum_{I} \overline{\varepsilon}_{I}\Big(
B^I\,[\chi,\overbar{B}_I]+
\,\varepsilon_I\,B^I\,\overbar{B}_I-\,M^I\,\overbar{M}_I
\Big)\bigg\}+O(x^{-3})~.
\notag
\end{align}
Notice that this quadratic action is invariant under the transformations (\ref{SU(4)})
even for $\varepsilon\not=0$. In other words, in the scaling limit we are considering,
the SU(4) symmetry of the initial action could be promoted to a U(4) symmetry.

The integrations over all moduli but $\chi$ are straightforward since all integrals are
Gaussian and can be readily performed. 
From the primed doublets $(\mu^\prime,h^\prime)$ and $(\overbar{\mu}^\prime,\overbar{h}^\prime)$ we obtain\,%
\footnote{Here and in the following we report the results of the integrations up to numerical factors that can always be reabsorbed in the overall normalization of the instanton partition function.}
\begin{equation}
\prod_{r=1}^M\det \big(\chi-m_r\big)_{\mathrm{fund}}
\label{int1}
\end{equation}
where $m_r$ are the eigenvalues of $m$ and the determinant is computed in the fundamental representation of U($k$). Integrating over the mixed fundamental doublets $(w,\mu)$ and $(\overbar{w},\overbar{\mu})$ yields
\begin{equation}
\prod_{u=1}^N\det\big(\chi-a_u\big)^{-1}_{\mathrm{fund}}
\label{int2}
\end{equation}
where $a_u$ are the eigenvalues of $a$. We observe that in order to get this result it is essential to have the
$\overline{a}$ parameter in the action which provides an effective ``mass''  for the fermions
$\mu$ and $\overbar{\mu}$. However, this $\overline{a}$ dependence cancels out in the final
result between the numerator and the denominator.

The integration over the adjoint doublet $(\lambda^{IJ},D^{IJ})$ gives
 \begin{equation}
\prod_{I<J=1}^3\det\big(\chi-\varepsilon_{IJ}\big)_{\mathrm{adj}}
\label{int3}
\end{equation}
while from the integration over $(B^I,M^I)$ and $(\overbar{B}_I,\overbar{M}_I)$ we get
\begin{equation}
\prod_{I=1}^4\det\big(\chi-\varepsilon_{I}\big)^{-1}_{\mathrm{adj}}
\label{int4}
\end{equation}
where the determinants are computed in the adjoint of U($k$).
In this case it is essential to have the $\overline{\varepsilon}_I$ terms
in the action since they provide a ``mass'' for the fermions $M^I$ and $\overbar{M}_I$. This 
is necessary to guarantee a non-vanishing result but, as before, the final answer 
(\ref{int4}) does not depend on these parameters.

The last integrations to be performed are those on the doublets $(\lambda,D)$
and $(\xi,\eta)$. One can explicitly verify that these Gaussian integrations yield just a
non-vanishing numerical constant. In fact, the entire dependence on $\chi$ and $\varepsilon$
cancels between the contribution of the fermions $\lambda$ and $\eta$ and the contribution
of the bosons $D$ and $\xi$. Again it is important in the intermediate steps to keep the
$\varepsilon$-dependence; in particular the term 
$\varepsilon\,\lambda\,\eta$ in the action acts as an effective ``mass''  for the components
of $\lambda$ and $\eta$ along the Cartan directions and it is crucial to have such a mass in order to
have a non-zero result. However, the final result does not depend 
on $\varepsilon$, which can therefore be safely set everywhere 
to zero at the end or kept at a non-zero value, if one wishes to do so.

Collecting all terms, we have
\begin{equation}
Z_k= \int d\chi~~ \frac{\prod_{r=1}^M\det \big(\chi-m_r\big)_{\mathrm{fund}}}
{\prod_{u=1}^N\det\big(\chi-a_u\big)_{\mathrm{fund}}}
~\frac{\prod_{I<J=1}^3\det\big(\chi-\varepsilon_{IJ}\big)_{\mathrm{adj}}}{\prod_{I=1}^4\det\big(\chi-\varepsilon_{I}\big)_{\mathrm{adj}}}~.
\label{Zk1}
\end{equation}
Exploiting the invariance of the integrand under U($k$) transformations, we can bring 
$\chi$ to the Cartan sub-algebra and, after introducing the Vandermonde determinant, 
integrate over the $k$ eigenvalues $\chi_i$ ($i=1,\ldots,k)$. 
Thus, we can rewrite (\ref{Zk1}) in the following way
\begin{equation}
Z_k=\int \prod_{i=1}^k d\chi_i~\prod_{i=1}^k
\frac{\prod_{r=1}^M(\chi_i-m_r)}{\prod_{u=1}^N(\chi_i-a_u)}\,
\prod_{i,j=1}^k 
\frac{\chi_{ij}^{1-\delta_{ij}}\,\prod_{I<J=1}^3(\chi_{ij}-\varepsilon_{IJ})}
{\prod_{I=1}^4(\chi_{ij}-\varepsilon_{I})}
\label{Zkfin}
\end{equation} 
with $\chi_{ij}=\chi_{i}-\chi_j$. 
For $\varepsilon=0$, the formula (\ref{Zkfin}) agrees with the one
proposed in \cite{Nekrasov:2018xsb} where it was conjectured to possibly derive from a
system with branes and anti-branes. This instanton partition function has been
recently considered also in \cite{Fucito:2020bjd,Bonelli:2020gku}.

From now on we set $\varepsilon=0$ and for simplicity present explicit results only for
$N=M$. In this case the instanton partition function $Z_k$ becomes 
dimensionless ($\chi_i$, $m_u$, $a_u$ and $\varepsilon_I$ have all 
dimensions of a mass) and the theory behaves as if it were conformal. 
With these positions we can rewrite (\ref{Zkfin}) as follows
\begin{equation}
Z_k=\mathcal{V}^{\,k}\int \prod_{i=1}^k d\chi_i~\prod_{i=1}^k\prod_{u=1}^N
\frac{(\chi_i-m_u)}{(\chi_i-a_u)}\,
\prod_{i<j=1}^k 
\frac{\chi_{ij}^{2}\,\prod_{I\not=J=1}^4(\chi_{ij}-\varepsilon_{IJ})}
{\prod_{I=1}^4(\chi_{ij}-\varepsilon_{I})}
\label{Zkfin2}
\end{equation} 
where
\begin{equation}
\mathcal{V}=-\frac{(\varepsilon_1+\varepsilon_2)(\varepsilon_1+\varepsilon_3)
(\varepsilon_2+\varepsilon_3)}
{\varepsilon_1\,\varepsilon_2\,\varepsilon_3\,\varepsilon_4}
\end{equation}
with $\varepsilon_4=-\varepsilon_1-\varepsilon_2-\varepsilon_3$.
Even if it is not at all obvious, we will show that $Z_k$ in (\ref{Zkfin2}) depends on $a_u$, $m_u$ 
and $\varepsilon_I$ only through the combination $\mathcal{V}\mathcal{M}$
where 
\begin{equation}
\mathcal{M}=\sum_{u=1}^N(m_u-a_u)~.
\label{Mis}
\end{equation}
Moreover, we will find that our explicit results for $Z_k$ are compatible with the small 
$q$ expansion of 
\begin{equation}
Z=\big[M(-q)\big]^{\mathcal{V}\mathcal{M}}
\label{Zmac}
\end{equation}
where 
\begin{equation}
M(q)=\prod_{n=1}^\infty\frac{1}{(1-q^n)^n}=1+q+3q^2+6q^3+13q^4+24q^5\ldots
\end{equation}
is the MacMahon function which counts the number of the planar partitions of the integers.
This result was also derived in \cite{Nekrasov:2017cih} using a conjectured 
plethystic exponential formula.

{From} (\ref{Zmac}) we may deduce that the non-perturbative prepotential is
\begin{equation}
\mathcal{F}=\varepsilon_1\varepsilon_2\varepsilon_3\varepsilon_4\,\log Z
=-(\varepsilon_1+\varepsilon_2)(\varepsilon_1+\varepsilon_3)(\varepsilon_2+\varepsilon_3)\,
\mathcal{M}\,\log M(-q)~.
\label{F8d}
\end{equation}

\subsection{The $k=1$ case}
At $k=1$ the instanton partition function is
\begin{equation}
Z_1=\mathcal{V}
\int\! d\chi~\prod_{u=1}^N\frac{(\chi-m_u)}{(\chi-a_u)}~.
\label{Z1inst}
\end{equation}
As usual, the integral over $\chi$ is computed by 
moving the poles $a_u$ in the upper-half complex plane and using the residue theorem
after closing the integration path at infinity \cite{Moore:1998et,Nekrasov:2002qd}. 
The result is simply
\begin{equation}
Z_1= \mathcal{V}\,\sum _{u=1}^N \frac{\prod_{v=1}^N (a_u-m_v)\phantom{\big|}}{
\prod_{v\not=u=1}^N(a_u-a_v)\phantom{\big|}}=-\mathcal{V}\mathcal{M}~.
\label{Z1fin}
\end{equation}

\subsection{Explicit results at higher instanton numbers}
To evaluate $Z_k$ at a generic instanton number $k$, we follow the prescription of 
\cite{Moore:1998et}. This corresponds to move again the poles $a_u$ in the upper-half complex 
plane, assign to each $\varepsilon_I$ a small imaginary part in such a way that
\begin{equation}
\mathrm{Im}(\varepsilon_4)\gg\mathrm{Im}(\varepsilon_3)\gg\mathrm{Im}(\varepsilon_2)\gg
\mathrm{Im}(\varepsilon_1)\gg\mathrm{Im}(a_u)~,
\label{prescription}
\end{equation}
and then integrate over each $\chi_i$ in lexicographic order by closing each 
contour in the upper-half plane. 
With this prescription\,%
\footnote{This is fully equivalent to the so-called Jeffrey-Kirwan prescription \cite{JK1995}.},
the poles that contribute to the result can be put in 
one-to-one correspondence with an $N$-array of four-dimensional Young tableaux $\{Y_u\}$ 
(with $u=1,\ldots,N$) containing a total number of $k$ boxes. Each box in the diagram $Y_u$
is labeled by four positive integers $(n_1,n_2,n_3,n_4)$ and is associated to a pole for one of
the integration variables in (\ref{Zkfin}) located at
\begin{equation}
\chi^{(n_1,n_2,n_3,n_4)}=
a_u+(n_1-1)\,\varepsilon_1+(n_2-1)\,\varepsilon_2+(n_3-1)\,\varepsilon_3+(n_4-1)\,\varepsilon_4~.
\end{equation}
For example, for $N=1$ and $k=2$ we have
\begin{equation}
Z_2=\mathcal{V}^2\!\int\!d\chi_1d\chi_2~
\,\frac{(\chi_1\!-\!m)(\chi_2\!-\!m)(\chi_1\!-\!\chi_2)^2\prod_{I\not= J=1}^4(\chi_{1}\!-\!\chi_2
\!-\!\varepsilon_{I}\!-\!\varepsilon_J)}{(\chi_1\!-\!a)(\chi_2\!-\!a)(\chi_{1}\!-\!\chi_2\!-\!\varepsilon_{1})
(\chi_{1}\!-\!\chi_2\!-\!\varepsilon_{2})(\chi_{1}\!-\!\chi_2\!-\!\varepsilon_{3})
(\chi_{1}\!-\!\chi_2\!-\!\varepsilon_{4})}
\end{equation}
and we see that the poles for $(\chi_1,\chi_2)$ are at
\begin{equation}
(a,a+\varepsilon_1)~,\quad(a,a+\varepsilon_2)~,\quad(a,a+\varepsilon_3)~,\quad
(a,a+\varepsilon_4)~,
\label{tableau1}
\end{equation}
which indeed correspond to the four four-dimensional Young tableaux with two boxes. 
These four tableaux are also associated to the four solid partitions of 2. This is a general feature
of the $N=1$ theory in which the poles that contribute to the integrals of $Z_k$ are 
in one-to-one correspondence with the solid partitions of $k$.

For $N=2$ and $k=2$, instead, the poles for $(\chi_1,\chi_2)$ turn out to be at
\begin{equation}
(a_1,a_2)~,\quad
(a_u,a_u+\varepsilon_1)~,\quad(a_u,a_u+\varepsilon_2)~,\quad(a_u,a_u+\varepsilon_3)~,\quad
(a_u,a_u+\varepsilon_4)
\label{tableau2}
\end{equation}
with $u=1,2$. The first pole location corresponds to a pair of tableaux with one box each; the other
locations describe pairs of tableaux, one with two boxes and the other with no boxes.
This example can be systematically generalized to other values of $N$ and $k$.

The evaluation of the residues is straightforward, even if tedious. 
When we sum up all contributions, remarkable simplifications occur and, as anticipated, 
the dependence on $a_u$, $m_u$ and $\varepsilon_I$ in the final result is 
only through the combination $\mathcal{V}\mathcal{M}$.
We have explicitly checked this fact up to $k=5$, and found
\begin{equation}
\begin{aligned}
Z_2&=\frac{1}{2}\,(\mathcal{V}\mathcal{M})^2+\frac{5}{2}\,\mathcal{V}\mathcal{M}~,\\
Z_3&=-\frac{1}{6}\,(\mathcal{V}\mathcal{M})^3-\frac{5}{2}\,(\mathcal{V}\mathcal{M})^2
-\frac{10}{3}\,\mathcal{V}\mathcal{M}~,\\
Z_4&=\frac{1}{24}\,(\mathcal{V}\mathcal{M})^4+\frac{5}{4}\,(\mathcal{V}\mathcal{M})^3
+\frac{155}{24}\,(\mathcal{V}\mathcal{M})^2+\frac{21}{4}\,\mathcal{V}\mathcal{M}~,\\
Z_5&=-\frac{1}{120}\,(\mathcal{V}\mathcal{M})^5-\frac{5}{12}\,(\mathcal{V}\mathcal{M})^4
-\frac{115}{24}\,(\mathcal{V}\mathcal{M})^3
-\frac{163}{12}\,(\mathcal{V}\mathcal{M})^2
-\frac{26}{5}\,\mathcal{V}\mathcal{M}~.
\end{aligned}
\label{Zkexpl}
\end{equation}
These results allow us to infer the formula (\ref{Zmac}).

Let us remark that, once individuated the poles through the prescription (\ref{prescription}), the evaluation of the residues does not introduce any sign ambiguity in summing up the contributions associated to the various partitions of the type considered in \cite{Nekrasov:2017cih,Nekrasov:2018xsb}. A particular sign choice is automatically picked up, and is exactly the one that leads to the correct exponentiation of the result. This choice is in agreement with the one pointed out in \cite{Fucito:2020bjd}. 

We mentioned above that the poles contributing to the integrals of $Z_k$ for the rank 1 theory are in 
one-to-one correspondence with the solid partitions of $k$. However, it is the generating function 
of the planar partitions $M(q)$ that appears in the end result (\ref{Zmac}). 
To understand this fact, let us specialize to $N=1$ with $a=0$ and $m=\varepsilon_4$, and then take the limit $\varepsilon_4\to0$. This limit clearly implies that $\mathcal{V}\mathcal{M}\to1$. 
It turns out that in this scaling all contributions to $Z_k$ corresponding to Young tableaux 
which contain at least one box with $n_4>1$ are set to zero, while all contributions associated 
to tableaux in which all boxes have $n_4=1$ are uniformly weighted with $(-1)^k$. 
In other words, in this limit the gran-canonical partition function (\ref{Zgrancanonical}) can be 
written as a weighted sum over the solid partitions $\pi$:
\begin{equation}
Z=\sum_{\pi}\mathsf{w}_{\pi}\,q^{|\pi|}
\label{weightedpartition}
\end{equation}
where $|\pi|$ is the size of $\pi$, and
\begin{equation}
\mathsf{w}_{\pi}=\begin{cases}
&(-1)^{|\pi|}\quad\mbox{if~$Y(\pi)$~does~not~extend~in~the~$\varepsilon_4$
\,direction}~,\\
&~~~~0\quad~~\,\,\mbox{~\,if~$Y(\pi)$~does~extend~in~the~$\varepsilon_4$
\,direction}~,
\end{cases}
\label{weight}
\end{equation}
with $Y(\pi)$ denoting the four-dimensional Young diagram associated to $\pi$. Given the form
of the weight function (\ref{weight}), it is clear that only the solid partitions $\pi$ which are also 
planar with respect to the $\varepsilon_4$-direction, contribute in the sum. Therefore, (\ref{weightedpartition}) actually coincides with the 
partition function of the planar partitions with uniform weight (up to an alternating sign), 
which are counted by the MacMahon function, {\it{i.e.}}
\begin{equation}
Z=M(-q)
\end{equation}
in agreement with (\ref{Zmac}) when $\mathcal{V}\mathcal{M}=1$. 
This argument can be extended to more general cases along the lines discussed in \cite{Cao:2017swr}.

\section{ADHM construction}
\label{secn:ADHM}
We have anticipated that the physical states of the open strings with at least one end-point on a 
D(--1)-brane can be put in correspondence with the moduli of instantonic configurations in eight dimensions. In this section we would like to elaborate on this. 

There are different definitions of eight-dimensional gauge instantons, which are briefly reviewed in a 
similar (although not identical) brane context in \cite{Billo':2009gc}. The definition appropriate to our 
current set-up is the one in which these configurations correspond to 
Yang Mills connections whose field strength satisfies the 
(anti) self-duality condition
\begin{equation}
{}^*\left(F \wedge F\right) =\pm \,F\wedge F~.   
	\label{self}
\end{equation}
When this condition holds, the quartic action $S_4$ for the gauge fields on the D7 branes, given in (\ref{action4}), drastically simplifies. Indeed, in this case
\begin{equation}
\tr \big(t_8\,F^4\big) = \pm\frac{1}{2} \tr (F\wedge F\wedge F \wedge F)~,
\end{equation}
and hence
\begin{equation}
S_4= \begin{cases}
-2\pi\ii\,  \overbar{\tau}\,k~,\\
-2\pi\ii\,\tau\,k~,
\end{cases}
\end{equation}
where $\tau$ is defined in (\ref{taudef}) and
\begin{equation}
k=\frac{1}{4!(2\pi)^4} \int \tr \left(F\wedge F\wedge F \wedge F\right)
\label{chern}
\end{equation}
is the fourth Chern number. This precisely matches the action of $k$ anti D-instantons or $k$ 
D-instantons, respectively.

In analogy with the familiar case of the four-dimensional instanton, one way to realize 
the conditions (\ref{self}) is to exploit the duality properties of the eight-dimensional Dirac 
matrices. Indeed, the chiral blocks $\sigma^{\mu\nu}$ 
(see (\ref{defsigmamunu})), satisfy
\begin{align}
	\label{dualsigma}
		\sigma^{[\mu_1\mu_2} \,\sigma^{\mu_3\mu_4]} 
		= +\frac{1}{4!}\,\epsilon^{\mu_1\mu_2\ldots \mu_7\mu_8}\,  \sigma_{\mu_5\mu_6}\, \sigma_{\mu_7\mu_8}
\end{align}
where in the left-hand side the indices are all anti-symmetrized, while the anti-chiral blocks
$\overbar{\sigma}^{\mu\nu}$ obey
\begin{align}
	\label{anti-dualsigma}
		\overbar{\sigma}^{[\mu_1\mu_2} \,\overbar{\sigma}^{\mu_3\mu_4]} 
		= -\frac{1}{4!}\,\epsilon^{\mu_1\mu_2\ldots \mu_7\mu_8}\, 
		\overbar{\sigma}_{\mu_5\mu_6}\, \overbar{\sigma}_{\mu_7\mu_8}~.
\end{align}
A field strength $F_{\mu\nu}$ proportional to $\sigma_{\mu\nu}$ or to 
$\overline{\sigma}_{\mu\nu}$ would thus enjoy the property (\ref{self}). 

To obtain such a field strength we can follow the ADHM construction in four dimensions
(for definiteness we choose the minus sign in (\ref{self})).
We first introduce the ADHM matrix\,%
\footnote{Since the construction involves matrices of various sizes, sometimes we find useful 
to indicate explicitly their sizes with the notation $A_{[n\times n]}$ for a matrix $A$ of size 
$n\times n$.}
\begin{equation}
\label{defDelta}
\Delta=
\begin{pmatrix}
\alpha_{[N \times 8k]}  \\[2mm]
(B-x)_{[8k \times 8k]} 
\end{pmatrix}
\end{equation}
where $\alpha$ is a $N\times 8k$ matrix, while $B$ and $x$ are $8k\times 8k$ matrices
given by 
\begin{align}
	\label{xmatis}
	B = (B_\mu)_{[k\times k]}\otimes  \overbar{\sigma}^\mu~,\qquad
	x=x_\mu \, \one_{[k \times k]}\otimes\overbar{\sigma}^\mu~.
\end{align} 
The matrices $\alpha$ and $B_\mu$ contain the bosonic moduli of the configuration,
while $x_\mu$ are the coordinates in eight dimensions.
Then, we define the gauge connection $A_\mu$, 
expressed as a $N\times N$ matrix, according to
\begin{equation}
A_\mu ={U}^\dagger \partial_\mu U~,
\label{con}
\end{equation}
where 
\begin{equation}
U=
\begin{pmatrix}
u_{[N \times N]}  \\
v_{[8k \times N]}     \\
\end{pmatrix}
\end{equation}
is a $(N+8k)\times N$ matrix subject to the following conditions
\begin{equation}
\Delta^\dagger U= U^\dagger\Delta=0~,
\quad
{U}^\dagger\, U=\mathbb{1}_{[N\times N]}~.
\label{du}
\end{equation}
The corresponding field strength is
\begin{equation}
\begin{aligned}
F_{\mu\nu} &= \partial_\mu A_\nu- \partial_\nu A_\mu +\big[A_\mu , A_\nu\big]
= 2\, \partial_{[\mu}  U^\dagger \,\partial_{\nu]} U+ 2\, U^\dagger \partial_{[\mu} U\,  {U}^\dagger \partial_{\nu]} U\\[1mm]
&=2 \,\partial_{[\mu}  {U}^\dagger  \big( \mathbb{1} - U {U}^\dagger    \big) 
\, \partial_{\nu]} U~.
\end{aligned}
\label{ff2}
\end{equation}
We now introduce a $8k\times 8k$ matrix $M$ such that 
\begin{equation}
\mathbb{1} - U {U}^\dagger =
\Delta \,M\,\Delta^\dagger  
\label{adhmguess0}
\end{equation}
or equivalently, thanks to the conditions (\ref{du}), such that
\begin{equation}
\Delta^\dagger \Delta =M^{-1}  ~.
\label{adhmguess}
\end{equation}
Plugging (\ref{adhmguess0}) into (\ref{ff2}) leads to
\begin{equation}
\begin{aligned}
F_{\mu\nu} &=2\, \partial_{[\mu} U^\dagger   \Delta \,M\, 
\Delta^\dagger \, \partial_{\nu]} U
= 2\,  U^\dagger\,  \partial_{[\mu}  \Delta\, M \, \partial_{\nu]} 
\Delta^\dagger \, U\\
&=2\,U^\dagger\begin{pmatrix}
0&0\\
0&\overbar{\sigma}_{[\mu}\, M\, \sigma_{\nu]}
\end{pmatrix} U = ~2\, \overbar{v}\,\overbar{\sigma}_{[\mu}\, M\, \sigma_{\nu]}\,v
\end{aligned}
\label{fmunu}
\end{equation}
where in the second line, with an abuse of notation, we have denoted $\mathbb{1}_{[k\times k]}\otimes (\overbar{\sigma}_\mu)_{[8\times 8]}$ and $\mathbb{1}_{[k\times k]}\otimes (\overbar{\sigma}_\mu)_{[8\times 8]}$ simply as $\overbar{\sigma}_\mu$ and $\sigma_\nu$.

If $N\geq 8$, it is possible to require that the $N\times N$ matrix $2\,\overbar{v}\,\overbar{\sigma}_{[\mu}\, M\, \sigma_{\nu]}\,v$ has a single non-trivial block of size $(8\times 8)$ proportional to 
$\overbar{\sigma}_{\mu\nu}$.
If this is the case, then, the field strength (\ref{fmunu}) 
satisfies the relation (\ref{self}) with the minus sign as a consequence
of the anti-self duality property (\ref{anti-dualsigma}).
Of course, this requirement puts stringent constraints on the form of the matrices $M$ and $\Delta$. 
For $k=1$, a solution is obtained by taking
\begin{equation}
\begin{aligned}
M&=f\,\mathbb{1}_{[8\times 8]}~,\quad
\alpha=\begin{pmatrix}
\rho\,\mathbb{1}_{[8\times 8]}\\
0_{\,[(N-8)\times 8]}
\end{pmatrix}~,\quad
\overbar{v}=f^{\frac{1}{2}}\,\alpha~,\\[1mm]
u&=f^{\frac{1}{2}}\begin{pmatrix}
-(B_\mu-x_\mu)\overbar{\sigma}^\mu&0_{[8\times (N-8)]}\\
0_{[(N-8)\times 8]}&0_{[(N-8)\times (N-8)]}
\end{pmatrix}
\end{aligned}
\label{ADHMdata}
\end{equation}
with $f=(\rho^2+r^2)^{-1}$ and $r^2=(x_\mu-B_\mu)(x^\mu-B^\mu)$. Through (\ref{fmunu}), 
this leads to the following field strength
\begin{equation}
F_{\mu\nu}=\frac{2\rho^2}{(\rho^2+r^2)^2}\begin{pmatrix}
\overbar{\sigma}_{\mu\nu}&0_{[8\times (N-8)]}\\
0_{[(N-8)\times 8]}&0_{[(N-8)\times (N-8)]}
\end{pmatrix}~.
\label{grossman}
\end{equation}
This represents an instanton solution of size $\rho$ and center $B_\mu$ which corresponds to
the embedding of the SO(8) octonionic
field strength of \cite{Grossman:1984pi,Grossman:1989bb} into U($N$) (with $N\geq 8$).
One can check that the gauge connection corresponding to (\ref{grossman}) is
\begin{equation}
A_{\mu}=\frac{(x^\nu-B^\nu)}{\rho^2+r^2}\begin{pmatrix}
\overbar{\sigma}_{\nu\mu}&0_{[8\times (N-8)]}\\
0_{[(N-8)\times 8]}&0_{[(N-8)\times (N-8)]}
\end{pmatrix}~,
\label{grossmanA}
\end{equation}
and that the fourth Chern number (\ref{chern}) is one  \cite{Grossman:1984pi,Grossman:1989bb}.
Generalizations to higher $k$ are possible.

In a supersymmetric theory, like the one we are considering, instantons preserve a 
fraction of supersymmetry. A supersymmetric instanton is then characterized
by the existence of a Killing spinor $\mathfrak{e}$ such that the 
supersymmetry variation of the gaugino
$\Lambda$ vanishes, namely
\begin{equation}
\delta\Lambda=\frac{1}{2}\,F_{\mu\nu}\,\gamma^{\mu\nu}\,\mathfrak{e}=0~,
\end{equation}
which reduces to
\begin{equation}
F_{\mu\nu}\,\sigma^{\mu\nu}\,\mathfrak{e} =0
\label{susy3}
\end{equation}
for a chiral $\mathfrak{e}$. In \cite{Bonelli:2020gku} it has been shown that to solve
this equation it is enough to require that
\begin{equation}
\Delta^\dagger \Delta \,\mathfrak{e}=f^{-1}_{[k\times k]} \otimes \mathfrak{e}
\label{adhm3}
\end{equation}
for some $(k\times k)$ matrix $f$. This means that the matrix $M$ introduced in (\ref{adhmguess})
must be such that $M\,\mathfrak{e}=f_{[k\times k]}\otimes \mathfrak{e}$.

To make contact with the previous sections, we write the matrices $B$ and $\alpha$ appearing in the 
ADHM matrix $\Delta$ in terms of the bosonic moduli corresponding to 
the open strings in the D(--1)/D7 brane system. In particular we take
the eight $k\times k$ matrices $B_\mu$, used to construct $B$ as in (\ref{xmatis}), 
to be given by the neutral moduli $B^I$ and 
$\overbar{B}_I$ of the D(--1)/D(--1) sector, and take the $N\times 8k$ matrix $\alpha$
to be related to the mixed moduli $\overbar{w}$ of the D7/D(--1) sector
according to $\alpha=\overbar{w}_{[N\times k]}\otimes \psi^\dagger_{[1\times 8]}$
where $\psi$ a reference Weyl chiral spinor \cite{Bonelli:2020gku}.
Given the choices we made in the string construction, we take 
\begin{equation}
\psi=\mathfrak{e}+\ii\,\mathfrak{e}^\prime
\end{equation}
with $\mathfrak{e}=\delta_{a,8}$ and $\mathfrak{e}^\prime=\delta_{a,7}$. The choice on the
Killing spinor $\mathfrak{e}$ reflects our choice of the preserved supersymmetry (and hence of the BRST charge) which, in turn, is related to how SO(8) has been broken to SO(7). The second choice on 
$\mathfrak{e}^\prime$ reflects the choice of the complex structure with which we have broken
SO(7) to SU(4) (see Appendix~\ref{app:notations}). With these positions it is easy to verify that
\begin{equation}
\psi\,\psi^\dagger\,\mathfrak{e}= -\tau^7\,\mathfrak{e}+\mathfrak{e}\quad\mbox{and}\quad
\sigma^{\mu\nu}\,\mathfrak{e}=-(\tau_m)^{\mu\nu}\,\tau^m\,\mathfrak{e}
\label{relations}
\end{equation}
where $\tau^m$ are the seven octonionic matrices defined in (\ref{mateltau}). Furthermore, we have
\begin{equation}
\begin{aligned}
\Delta^\dagger\Delta&=\alpha^\dagger\alpha+(B^\dagger-x^\dagger)(B-x)\\
&=
w\overbar{w}\otimes
\psi\psi^\dagger+\frac{1}{2}\,\big[B_\mu\,,\,B_\nu]\otimes\sigma^{\mu\nu}+
(B_\mu-x_\mu)(B^\mu-x^\mu)\otimes \mathbb{1}_{[8\times 8]}~,
\end{aligned}
\end{equation}
and thus projecting onto $\mathfrak{e}$ and using the relations (\ref{relations}), we deduce that
in order to satisfy (\ref{adhm3}) one must require that
\begin{equation}
\frac{1}{2}\,\big[B_\mu\,,\,B_\nu](\tau_m)^{\mu\nu}+w\overbar{w}\,\delta_m^7=0~.
\end{equation}
In complex notation, these equations read
\begin{equation}
\big[B^I,\overbar{B}_I\big]+w\,\overbar{w}=0~,\qquad
\big[\overbar{B}_I,\overbar{B}_J\big]+\frac{1}{2}\,\epsilon_{IJKL}\,\big[B^K,B^L\big]=0~,
\label{bosconstraints1}
\end{equation} 
which are exactly the constraints reported in (\ref{bosconstraints}) and derived from the moduli action in the D(--1)/D7-brane system.

We observe that for $k=1$ the first constraint leads to $w=\overbar{w}=0$. This implies that 
the ADHM data reduce to those in (\ref{ADHMdata}) for $\rho=0$, thus describing 
a point-like instanton configuration with vanishing size. It would be interesting to explore what happens if one introduces a non-commutative deformation parameter in such a way that the 
right-hand side of the real constraint in (\ref{bosconstraints1}) differs from zero, thus 
allowing to have configurations with non-zero size, and more generally to investigate the features of
the solutions of the ADHM constraints in the general case. We leave these issues to future 
investigations.

\section{Conclusions}
\label{secn:conclusions}

In this paper we have studied a D(--1)/D7-brane system in Type IIB string theory that
describes the non-perturbative sector of a U($N$) gauge theory in eight dimensions, 
and provided an explicit analysis of the massless open string states and of their vertex operators
in the various sectors. The new ingredient with respect to previous work is the
introduction of a constant magnetic flux on the world-volume of the D7-branes which allows
for the existence of bosonic moduli in the mixed sectors.
After discussing the moduli action, we have computed the instanton partition function
using localization and confirmed the results of \cite{Nekrasov:2017cih} obtaining 
a closed form expression in terms of the MacMahon function.

It is important to remark that this instanton partition function and the corresponding prepotential
given in (\ref{F8d}) involve only the fields of the overall U(1) factor inside U($N$). Furthermore, the prepotential is cubic in the $\varepsilon_I$ parameters used in the localization process, showing
that the gauge effective action receives non-perturbative corrections only in 
curved space. To see this in an explicit way, let us first turn off the parameters 
$m_u$ so that the quantity $\mathcal{M}$ in (\ref{Mis}) simply reduces to $\tr \phi$, where $\phi$ is the scalar component of the
vector superfield (see (\ref{Phi})). Then let us note that 
\begin{equation}
(\varepsilon_1+\varepsilon_2)(\varepsilon_1+\varepsilon_3)(\varepsilon_2+\varepsilon_3)
\,\sim \,t_8^{\mu_1\ldots\mu_8}\,\mathcal{W}_{\mu_1\mu_2}\,\mathcal{W}_{\mu_3\mu_4}
\,\mathcal{W}_{\mu_5\mu_6}\,\mathcal{I}_{\mu_7\mu_8}
\label{e3}
\end{equation}
where $\mathcal{W}_{\mu\nu}\,\equiv\,\mathcal{W}_{\mu\nu z}$ is the graviphoton field strength defined in (\ref{W}). The tensor $\mathcal{I}_{\mu\nu}$, instead, describes the background magnetic 
flux on the D7-branes and is obtained from (\ref{background}) by setting $f_I=-1$ for all $I=1,\ldots 4$\,%
\footnote{According to (\ref{thetaf}), this corresponds to the choice of the twist parameters $\theta_I = 1/4$ that we have made in (\ref{theta14}).}. Finally,
$t_8$ is the tensor that typically appears in higher derivative actions, like for instance 
(\ref{action4})\,%
\footnote{Given the particular form of $\mathcal{W}_{\mu\nu}$ and $\mathcal{I}_{\mu\nu}$, one
has
\begin{equation*}
t_8^{\mu_1\ldots\mu_8}\,\mathcal{W}_{\mu_1\mu_2}\,\mathcal{W}_{\mu_3\mu_4}
\,\mathcal{W}_{\mu_5\mu_6}\,\mathcal{I}_{\mu_7\mu_8} \sim 
2 \,
\tr\big(\mathcal{W}\,\mathcal{W}\,\mathcal{W}\,
\mathcal{I}\big)-\frac{1}{2}\,
\tr\big(\mathcal{W}\,\mathcal{W} \big)\,
\tr\big(\mathcal{W}\,\mathcal{I} \big)
\end{equation*} 
where the trace is over the space-time indices, from which the result (\ref{e3}) easily follows.}. Suppressing all indices for simplicity, the prepotential (\ref{F8d}) is therefore
\begin{equation}
\mathcal{F}\,\sim \,t_8 \,\mathcal{W}^3\,\mathcal{I}\,\tr \phi \,\log M(-q)~.
\label{Fconcl}
\end{equation}
To obtain the effective action, we promote $\phi$ to the full scalar superfield $\Phi(x,\theta)$
given in (\ref{Phi}) and the graviphoton field strength $\mathcal{W}_{\mu\nu}$ 
to the full graviphoton superfield \cite{Billo:2006jm} 
\begin{equation}
W_{\mu\nu}(x,\theta)=\mathcal{W}_{\mu\nu}(x)+\theta\,\chi_{\mu\nu}(x)
+\frac{1}{2}\,\theta\sigma^{\lambda\rho}\,\theta\, R_{\mu\nu\lambda\rho}(x)+\ldots
\end{equation}
where $\chi_{\mu\nu}(x)$ and $R_{\mu\nu\rho\sigma}(x)$ are respectively, the
self-dual parts of the gravitino field strength and of the Riemann tensor.
In this way we obtain a prepotential $\mathcal{F}(\Phi,W)$ that is linear 
in $\Phi$ and cubic in $W$. The eight-dimensional effective action is then given by
\begin{equation}
\int d^8x \,d^8 \theta \,\mathcal{ F}(\Phi, W) +\mathrm{c.c.}
\end{equation}
After integration over the $\theta$'s, our results predict an all-instanton formula for higher derivative couplings in Type II B supergravity, which are schematically of the form $t_8\,t_8\, R^3\tr F\,F_{\mathrm{class}}$, in presence of D7-branes carrying a non-trivial world-volume flux
$F_{\mathrm{class}}\sim \mathcal{I}$ along a U(1) subgroup of 
U($N$). It would be very interesting to further investigate the properties and the implications of 
these non-perturbative 
gravitational couplings, and also to explore the structure of the instanton partition 
function of our system when the graviphoton parameters $\varepsilon_I$ do not add up to zero.

\vskip 1.5cm
\noindent {\large {\bf Acknowledgments}}
\vskip 0.2cm
We would like to thank Igor Pesando for helpful discussions.
This research is partly supported by the INFN Iniziativa Specifica ST\&FI
``String Theory \& Fundamental Interactions''.
\vskip 1cm
\begin{appendix}
\section{Notations and conventions}
\label{app:notations}
In this appendix we collect our notations for the various symmetry groups of the brane system and
for the spinors and Dirac matrices.

\subsection*{From ten to eight dimensions}
\label{app:sptsymm}
The symmetry of the ten-dimensional flat background in which the superstring theory is defined is broken by the presence of the seven-branes as 
\begin{align}
	\label{10to82}
		\mathrm{SO}(10)\to \mathrm{SO}(8) \otimes \mathrm{SO}(2)~,
\end{align}
with $\mathrm{SO}(8)$ rotating the first eight coordinates $x^\mu$. Thus, 
the vector representation decomposes as
\begin{align}
	\label{10vdec}
		\mathbf{10}\to (\mathbf{8}_v,0) \oplus (\mathbf{1},+1) \oplus (\mathbf{1},-1)~,
\end{align}
where each couple $(\mathbf{R},p)$ indicates a representation $\mathbf{R}$ of $\mathrm{SO}(8)$ and a charge $p$ under $\mathrm{U}(1)\sim \mathrm{SO}(2)$. The  
$\mathrm{U}(1)^4$ Cartan subgroup of $\mathrm{SO}(8)$ corresponds to the rotations in the 
$(x^{2I-1},x^{2I})$ planes, with $I=1, \ldots 4$, and here we will denote its four parameters 
as $\varepsilon_I$.

Under the breaking (\ref{10to82}), the 32-components of the ten dimensional spinors 
(which can be chiral or anti-chiral) decompose into $S^A S^\pm$, where $S^A$ are the 
16 components of $\mathrm{SO}(8)$ spinors (chiral or anti-chiral)
and $S^\pm$ have charge $\pm 1/2$ with respect to $\mathrm{U}(1)$. 
More precisely, we have
\begin{align}
	\label{spin1082}
		\mathbf{16}_s & \to\big (\mathbf{8}_s,+1/2\big) \oplus \big(\mathbf{8}_c,-1/2\big)~,
		\notag\\
		\mathbf{16}_c & \to \big(\mathbf{8}_s,-1/2\big) \oplus \big(\mathbf{8}_c,+1/2\big)~.
\end{align}
The ten-dimensional Dirac matrices take the form
\begin{align}
\label{10dgamma}
\Gamma^\mu = \gamma^\mu \otimes \mathbb{1}~,~~~
\Gamma^9 = \gamma\otimes \sigma_1~,~~~
\Gamma^{10} = \gamma\otimes \sigma_2~,
\end{align} 
where $\gamma^\mu$ with $\mu=1,2\ldots 8$ are the eight-dimensional Dirac matrices, 
$\gamma$ is the eight-dimensional chirality matrix and the sigmas are the usual Pauli matrices.
The ten-dimensional charge conjugation matrix $\mathcal{C}$ satisfying $\mathcal{C} \,\Gamma^M \left(\mathcal{C}^{-1}\right)^T = -\left(\Gamma^M\right)^T$ for $M=1,\ldots,10$ 
is written as
\begin{align}
\label{C10toC8}
\mathcal{C} = C \otimes \sigma_2~, 
\end{align}      
where $C$ is the eight-dimensional charge conjugation which is such that
$C \,\gamma^\mu \left(C^{-1}\right)^T = -\left(\gamma^\mu\right)^T$.

\subsection*{Eight-dimensional spinors}
\label{app:spinors}
For the eight-dimensional spinors we can use either the ``spin field'' basis or the Majorana-Weyl
basis, which we are going to describe.
\paragraph{$\bullet$ The spin field basis:}
In the  ``spin field'' basis the spinor components are labeled by (twice) their SO(8) weight vectors
and are 
\begin{equation}
\begin{aligned}
		S^A & = \Big(S^{++++}\,,\,S^{-+++}\,,\,S^{+-++}\,,\,S^{--++}\,,\,S^{++-+}\,,\,
		S^{-+-+}\,,\,S^{+--+}\,,\,S^{---+}\,,
		\\
		& ~~~~~~
		S^{+++-}\,,\,S^{-++-}\,,\,S^{+-+-}\,,\,S^{--+-}\,,\,S^{++--}\,,\,S^{-+--}\,,\,S^{+---}\,,\,S^{----}\Big)~.
\end{aligned}	
\label{sfb}
\end{equation}
In this basis the Dirac matrices are 
\begin{equation}
\begin{aligned}
		\gamma^1 & = \sigma_1\otimes \mathbb{1}\otimes \mathbb{1} \otimes \mathbb{1}~,~~~~~~
		\gamma^2 = \sigma_2\otimes \mathbb{1}\otimes \mathbb{1} \otimes \mathbb{1}~,\\
		\gamma^3 & = \sigma_3\otimes \sigma_1\otimes \mathbb{1} \otimes \mathbb{1}~,~~~~~
		\gamma^4 = \sigma_3\otimes \sigma_2\otimes \mathbb{1} \otimes \mathbb{1}~,\\
		\gamma^5 & = \sigma_3\otimes \sigma_3\otimes \sigma_1 \otimes \mathbb{1}~,~~~\,
		\gamma^6 = \sigma_3\otimes \sigma_3\otimes \sigma_2 \otimes \mathbb{1}~,\\
		\gamma^7 & = \sigma_3\otimes \sigma_3\otimes \sigma_3 \otimes \sigma_1~,~~\,
		\gamma^8 = \sigma_3\otimes \sigma_3\otimes \sigma_3 \otimes \sigma_2~.
\end{aligned}
\label{g8dsfb}
\end{equation}
The chirality matrix is
\begin{equation}
	\label{chir8sfb}
		\gamma = \gamma^1 \gamma^2 \gamma^3 \gamma^4 \gamma^5 \gamma^6 \gamma^7 \gamma^8
		= \sigma_3 \otimes \sigma_3\otimes \sigma_3 \otimes \sigma_3~,  
\end{equation}
while the charge conjugation matrix reads
\begin{equation}
	\label{Cp8}
		C = \sigma_2 \otimes \sigma_1 \otimes \sigma_2 \otimes \sigma_1~.
\end{equation}
The chirality matrix is diagonal, and the eight components in (\ref{sfb}) with an even number of minuses are chiral, {\it{i.e.}} they span the $\mathbf{8}_s$ representation of SO(8), while 
the eight ones with an odd number of minuses are anti-chiral,  {\it{i.e.}} they  span 
the $\mathbf{8}_c$ representation.   

\paragraph{$\bullet$ The Majorana-Weyl basis:}
For $\mathrm{SO}(8)$ it is also possible to use a Majorana-Weyl basis. This can be achieved introducing the chiral combinations
\begin{align}
	\label{Scis}
		\cS^\alpha 
		= \frac{1}{\sqrt{2}} 
		\begin{pmatrix}
				\ii \,S^{+--+}- \ii\,S^{-++-}\\[1mm]
				- S^{+--+}- S^{-++-} \\[1mm]
				-\ii\,S^{+-+-}-\ii\,S^{-+-+}\\[1mm]
				-S^{+-+-}+S^{-+-+}\\[1mm]
				-\ii \,S^{++--}+\ii\,S^{--++}\\[1mm]
				-S^{++--}-S^{--++}\\[1mm]
				-\ii\,S^{++++}-\ii\,S^{----}\\[1mm]
				S^{++++} - S^{----}
		\end{pmatrix}~,
\end{align}
and the anti-chiral ones
\begin{align}
	\label{Sacis}
		\cS^{\dot{\alpha}} 
		= \frac{1}{\sqrt{2}} 
		\begin{pmatrix}
			\ii\,S^{+---}-\ii\, S^{-+++}\\[1mm]
			-S^{+---}- S^{-+++}\\[1mm]
			-\ii\,S^{+-++}-\ii\, S^{-+--}\\[1mm] 
			-S^{+-++}+S^{-+--}\\[1mm]
			-\ii\,S^{++-+}+\ii\,S^{--+-}\\[1mm]
			-S^{++-+}- S^{--+-}\\[1mm]
			-\ii\,S^{+++-}-\ii\,S^{---+}\\[1mm]
			S^{+++-} - S^{---+}
		\end{pmatrix}~.
\end{align}
In the basis $(\cS^\alpha,\cS^{\dot{\alpha}})$, the chirality matrix is $\mathbb{1}_8\otimes \sigma_3$, the charge conjugation is the identity and all Dirac matrices are purely imaginary. 
They take the form\,%
\footnote{With an abuse of notation we denote these matrices again by $\gamma^\mu$, even if they do not coincide with the ones in (\ref{g8dsfb}); rather they are equivalent to the latter under the change 
of basis in (\ref{Scis}) and (\ref{Sacis}).}
\begin{align}
	\label{defg8}
		{\gamma}^m = \tau^m\otimes\sigma_1~,~~~\gamma^8 = \mathbb{1}_8\otimes \sigma_2~,
\end{align}
where $\tau^m$, with $m=1,\ldots, 7$, are ($-\ii$ times) seven-dimensional Dirac matrices. 
These latter have elements
\begin{align}
	\label{mateltau}
		\left(\tau^m\right)_{\alpha\beta} & = - \ii \left(\delta^{m}_{\alpha}\,\delta^{8}_\beta
		-\delta^{m}_\beta\,\delta^8_\alpha + c_{m\alpha\beta}\right)~,
\end{align}
where the completely anti-symmetric tensor $c_{mnp}$ describes the octonionic structure constants:
\begin{align}
\label{octc}
c_{127} = c_{163} = c_{154} = c_{253} = c_{246} = c_{347} = c_{567} = 1~,
\end{align}
with all other  elements being zero. The Dirac matrices in (\ref{defg8}) can also be written as
\begin{align}
	\label{gammasigma}
		\gamma^\mu =
		\begin{pmatrix}
			0 & \sigma^\mu \cr
			\overbar{\sigma}^\mu & 0  
		\end{pmatrix}~,
\end{align}
with $\sigma^\mu = \big(\tau^m,-\ii\,\mathbb{1}\big)$ and $\overbar{\sigma}^\mu = 
\big(\tau^m,\ii\,\mathbb{1}\big)$. 
We also write 
\begin{align}
	\label{defsigmamunu}
		\gamma^{\mu\nu} = \frac 12 \big[\gamma^\mu\,,\,\gamma^\nu\big] =
		\begin{pmatrix}
			\sigma^{\mu\nu} & 0 \cr
			0 & \overbar{\sigma}^{\mu\nu}   
		\end{pmatrix}~,
\end{align}
with
\begin{align}
	\label{smunu}
		\sigma^{mn} & = \overbar{\sigma}^{mn} = \tau^{mn} = \frac 12 \big[\tau^m\,,\,\tau^n\big]
		~,~~~\sigma^{m8} = - \overbar{\sigma}^{m8} = \ii \,\tau^m~. 
\end{align}
This Majorana-Weyl basis is the one used in \cite{Billo:2009di}. Using (\ref{Scis}) and (\ref{Sacis}) 
we can rewrite all quantities involving fermions given there, in the spin field basis used in this paper.

\subsection*{The $\mathrm{U}(4)$ symmetry}
\label{subsec:U4}
The  SO(8) symmetry contains a U(4) subgroup that preserves the complex structure introduced in (\ref{complexstructure}). The U(1)$^4$ Cartan subgroup of SO(8) in in U(4) and
acts on the complex coordinates as
\begin{align}
	\label{Cartan4}
		z^I \to \rme^{\ii\,\varepsilon_I} z^I~.
\end{align}
One has
$\mathrm{U}(4)\simeq \big(\mathrm{SU}(4)\otimes \mathrm{U}(1)\big)/\mathbb{Z}_4$, 
where $U(1)$ is the diagonal action in the U(1)$^4$ subgroup parametrized by 
$\varepsilon/4$, where
\begin{align}
	\label{defvare}
		\varepsilon = \varepsilon_1+\varepsilon_2+\varepsilon_3+\varepsilon_4~.
\end{align}
The Cartan subgroup of SU(4) is spanned by the ``traceless'' parameters
$\widehat{\varepsilon}_I = \varepsilon_I - \varepsilon/4$, out of which only three are independent. 
We can denote the U(4) representations by $\mathbf{r}_q$, where $\mathbf{r}$ is a 
SU(4) representation and $q$ a U(1) charge. With respect to U(4), the relevant SO(8) 
representations decompose according to the following pattern. 
For the vector we have\,%
\footnote{We fix the convention on the U(1) charge by declaring that the charge of the fundamental representation, whose states rotate by $\exp(\ii\,\varepsilon/4)$, is 1.} 
\begin{align}
	\label{8vu4}
		\mathbf{8}_v \to \mathbf{4}_1 \oplus \mathbf{\overbar{4}}_{-1}~,	
\end{align}
the two representations in the right-hand side corresponding to $z^I$ and $\bar z^I$. 
For the chiral spinor we have
\begin{align}
	\label{8su4}
		\mathbf{8}_s \to \mathbf{1}_{2} \oplus \mathbf{6}_0 \oplus \mathbf{1}_{-2}~.
\end{align}
The two SU(4) singlets are, respectively,\,%
\footnote{The minus signs appearing in (\ref{SbarS}) and later in (\ref{SIJ}) and (\ref{SI}) are inserted to be consistent with the action of the charge conjugation matrix (\ref{Cp8}) 
on the spin fields.} 
\begin{equation}
	\label{SbarS}
		\overbar{S} = S^{++++}~,\qquad S = - S^{----}~,
\end{equation}
while the states of the $\mathbf{6}_0$ representation are $S^{IJ}$ with $I,J=1,\ldots 4$ and $S^{IJ} = - S^{JI}$. Their identification in the spin field basis is
\begin{align}
	\label{SIJ}	
		S^{12} & = S^{++--}~,~~~
		S^{13} = -S^{+-+-}~,~~~
		S^{14} = S^{+--+}~,
		\notag\\
		S^{23} & = S^{-++-}~,~~~
		S^{24} = -S^{-+-+}~,~~~
		S^{34} = S^{--++}~.
\end{align}
Thus $S^{IJ}$ is proportional to the spin field component which has a plus in the
$I$-th and $J$-th directions. The conjugates $\overbar{S}_{IJ}$, which are the spin fields with a minus in the $I$-th and $J$-th directions, are not independent. Indeed we have
\begin{equation}
	\label{SIJtoSbIJ}
		\overbar{S}_{IJ}=\frac{1}{2}\,\epsilon_{IJKL}\,S^{KL}~.
\end{equation}
The anti-chiral spinor representation decomposes as
\begin{align}
	\label{antidec}
		\mathbf{8}_c \to \mathbf{\overbar{4}}_{1} \oplus \mathbf{4}_{-1} ~.
\end{align}
The states in the two representations in the right-hand side are respectively given by
\begin{equation}
\overbar{S}_I=\begin{pmatrix}
               \displaystyle{S^{-+++}}
               \\[1mm]
               \displaystyle{S^{+-++}}
               \\[1mm]
               \displaystyle{S^{++-+}}
               \\[1mm]
               \displaystyle{S^{+++-}}
            \end{pmatrix}~,\qquad
S^I=\begin{pmatrix}
               \displaystyle{S^{+---}}
               \\[1mm]
               \displaystyle{-S^{-+--}}
               \\[1mm]
               \displaystyle{S^{--+-}}
               \\[1mm]
               \displaystyle{-S^{---+}}
            \end{pmatrix}~.
\label{SI}
\end{equation}
Using first the change of basis (\ref{Scis}) and (\ref{Sacis}), and then the identifications (\ref{SbarS}), 
(\ref{SIJ}) and (\ref{SI}) we can express all quantities involving the fermions given in \cite{Billo:2009di} 
in terms of the U(4) notation used in the main text.

\subsection*{The $\mathrm{SO}(7)$ symmetry}
The moduli action described in Section~\ref{secn:moduliaction} possesses a BRST charge which in 
our conventions is the last component $\cS^8$ of a chiral spinor
(see (\ref{BRST}) and (\ref{QbarQ})). 
The SO(8) subgroup that preserves the BRST charge is SO(7), embedded in a non-standard fashion\,%
\footnote{This embedding is exchanged by triality with the standard one in which $\mathbf{8}_v\to \mathbf{7} \oplus \mathbf{1}$ or with the embedding where it is the anti-chiral spinor that decomposes in such a way.} in which
the chiral spinor decomposes as $\mathbf{8}_s \to \mathbf{7} \oplus \mathbf{1}$, 
where $\mathbf{7}$ is the vector representation of $\mathrm{SO}(7)$. More explicitly, we
have $\cS^\alpha \to (\cS^m,\cS^8)$ with $m=1,\ldots,7$.
Both the anti-chiral spinor $\mathbf{8}_c$ and the vector\,%
\footnote{To be precise, given the vector $v^\mu$, it is $(v^m,-v^8)$ that
transforms as a spinor.} $\mathbf{8}_v$ become spinors of SO(7). 
The adjoint representation of SO(8) decomposes according to 
$\mathbf{28} \to \mathbf{21} + \mathbf{7}$, where $\mathbf{21}$ is the adjoint of 
SO(7). In particular, this means that an anti-symmetric tensor of SO(8), like for example 
$D_ {\mu\nu}$, decomposes into
\begin{align}
	\label{W21}
		D_{\mu\nu}^{\mathbf{21}} =\frac{1}{2}\, D_{mn} \left(\tau^{mn}\right)_{\mu\nu}~,~~~
		D_{\mu\nu}^{\mathbf{7}} = D_{m} \left(\tau^{m}\right)_{\mu\nu}~.
\end{align}  
Here $D_{mn}$ and $D_{m}$ are, respectively, an anti-symmetric tensor and a vector
of $\mathrm{SO}(7)$, while $\tau^m$ and $\tau^{mn}$ are the matrices introduced in (\ref{mateltau}) and (\ref{smunu}). 

This is another way in which the SO(7) symmetry emerges in our brane system and in the ADHM construction described in section \ref{secn:ADHM}; indeed, to disentangle the quartic term proportional to $\comm{a^\mu}{a^\nu} \comm{a_\mu}{a_\nu}$ 
in the  $\mathrm{U}(k)$ moduli action, thanks to the Jacobi identity, the minimal required set of auxiliary fields is not a generic anti-symmetric tensor $D_{\mu\nu}$ but its $\mathbf{7}$ part only.        
The $\mathrm{U}(1)^3$ Cartan subgroup of SO(7) is embedded into the Cartan 
subgroup of SO(8) by restricting the four parameters $\epsilon_I$ to be traceless, {\it{i.e.}} 
to satisfy the requirement $\varepsilon = 0$. Only in this case is the spinor $\cS^8$ invariant. 
Note that this is the same condition that defines the Cartan subgroup of the $\mathrm{SU}(4)\subset \mathrm{U(4)}$ symmetry.   

\subsection*{The $\mathrm{SO}(6)\simeq\mathrm{SU}(4)$ symmetry}
\label{subsec:SO6Su4}
The subgroup of SO(7) that is compatible with the complex structure  
(\ref{complexstructure}) is the SO(6) group under which the SO(7) vector decomposes 
as $\mathbf{7}\to \mathbf{6} \oplus \mathbf{1}$, namely $V^m \to (V^{\widehat{m}},V^7)$, with $\widehat{m}=1,\ldots 6$ and the spinor representation of SO(7) decomposes as $\mathbf{8}\to \mathbf{4}_s \oplus \mathbf{4}_c$, the two addends being the chiral and anti-chiral spinors 
of SO(6). The Cartan subgroup of this SO(6) coincides with the one of SO(7). 

This SO(6) group is isomorphic to the SU(4) subgroup of U(4) discussed above. Under this isomorphism, the chiral and anti-chiral spinors are mapped to the fundamental and anti-fundamental representations $\mathbf{4}$ and $\bar{\mathbf{4}}$, while a vector like $D^{\widehat{m}}$ 
is mapped to the anti-symmetric tensor $D^{IJ}$, with $I,J=1,\ldots 4$, satisfying the
constraint (\ref{SIJtoSbIJ}).

\section{Technical details}
\label{app:details}
In this appendix we collect some technical material related to the string theory and conformal field theory methods used in
the main text.
\subsection*{Twisted coordinates and mode expansions}
\label{app_twist}
In \cite{Bertolini:2005qh} the conformal field theory of bosonic and fermionic fields
with non-trivial monodromy properties has been described in full detail.
Here we make use of the results of that reference and adapt them to the case of interest.
The complex bosonic and fermionic string coordinates along the longitudinal directions in the
D7/D7$^\prime$ sectors satisfy the monodromy properties (\ref{monodromyZ}) and 
(\ref{monodromyPsi}) around the origin of the world-sheet.
After canonical quantization, these fields have the following mode expansions
\begin{equation}
\begin{aligned}
\ii\,\partial Z^I(z)&=\sqrt{2\alpha^\prime}\,\Big(\sum_{n=1}^\infty 
\overbar{a}_{n-\theta_I}^I\,z^{-n+\theta_I-1}+\sum_{n=0}^\infty 
{a}_{n+\theta_I}^{\dagger\,I}\,z^{n+\theta_I-1}\Big)~,\\
\ii\,\partial \overbar{Z}_I(z)&=\sqrt{2\alpha^\prime}\Big(\sum_{n=0}^\infty 
{a}_{I,\,n+\theta_I}\,z^{-n-\theta_I-1}+\sum_{n=1}^\infty 
\overbar{a}_{I,\,n-\theta_I}^{\dagger}\,z^{n-\theta_I-1}\Big)~,
\end{aligned}
\label{ZbarZexp}
\end{equation} 
and
\begin{equation}
\begin{aligned}
\Psi^I(z)&=\sqrt{2\alpha^\prime}\sum_{n=0+\nu}^\infty \Big(\,
\overbar{\Psi}_{n-\theta_I}^I\,z^{-n+\theta_I-\frac{1}{2}}+ 
{\Psi}_{n+\theta_I}^{\dagger\,I}\,z^{n+\theta_I-\frac{1}{2}}\,\Big)~,\\
\overbar{\Psi}_I(z)&=\sqrt{2\alpha^\prime}\sum_{n=0+\nu}^\infty 
\Big(\,
{\Psi}_{I,\,n+\theta_I}\,z^{-n-\theta_I-\frac{1}{2}}+\sum_{n=0+\nu}^\infty 
\overbar{\Psi}_{I,\,n-\theta_I}^{\dagger}\,z^{n-\theta_I-\frac{1}{2}}\Big)
\end{aligned}
\label{PsibarPsiexp}
\end{equation}
for the fermionic coordinates with $\nu=0$ in the R sector and $\nu=\frac{1}{2}$ in the 
NS sector. The bosonic oscillators satisfy the canonical commutation relations
\begin{equation}
\begin{aligned}
\Big[a_{I,\,n+\theta_I},a^{\dagger\, J}_{m+\theta_J}\Big]&=(n+\theta_I)\,\delta_I^{J}\,
\delta_{nm}\quad\quad \forall\, n,m\ge 0~, \\[1mm]
\Big[\overbar{a}^I_{n-\theta_I},\overbar{a}^{\dagger}_{J,\,m-\theta_J}\Big]&=
(n-\theta_I)\,\delta^{I}_J\,\delta_{nm}\quad\quad \forall\, n,m\ge 1~,
\end{aligned}
\end{equation}
while the fermionic oscillators obey the canonical anti-commutators
\begin{equation}
\begin{aligned}
\Big\{\Psi_{I,\,n+\theta_I},\Psi^{\dagger\, J}_{m+\theta_J}\Big\}&=
\Big\{\overbar{\Psi}^J_{n-\theta_J},\overbar{\Psi}^{\dagger}_{I,\,m-\theta_I}\Big\}
=\delta_I^{J}\,\delta_{nm}\quad\quad
 \forall \,n,m\ge 0+\nu~.
\end{aligned}
\end{equation}
Notice that if all $\theta_I$'s are different from zero (which is the case
considered in the main text), there are no bosonic oscillators with index 0, and hence the 
momentum can not be defined in any of these directions.

Following \cite{Bertolini:2005qh}, we define the Virasoro generators 
$L_n$ for the D7/D7$^\prime$ strings. In particular we find that the bosonic 
contribution to $L_0$ is 
\begin{equation}
L_0^{(Z)}=\widehat{N}^{(Z)}+ \frac{1}{2}\sum_{I=1}^4\theta_I(1-\theta_I)
\label{L0Z}
\end{equation}
where $\widehat{N}^{(Z)}$ is the bosonic number operator
\begin{equation}
\widehat{N}^{(Z)}=\sum_{I=1}^4\bigg[\sum_{n=1}^\infty
\overbar{a}^{\dagger}_{I,\,n-\theta_I}\overbar{a}_{n-\theta_I}^I
+\sum_{n=0}^\infty a^{\dagger\,I}_{n+\theta_I}a_{I,\,n+\theta_I}\bigg]+
\sum_{n=1}^\infty \big(
\overbar{a}^{\dagger}_{5,\,n}\overbar{a}_{n}^5
+a^{\dagger\,5}_{n}a_{5,\,n}\big)~,
\label{NZ}
\end{equation}
while the fermionic contribution to $L_0$ is
\begin{equation}
L_0^{(\Psi)}=\begin{cases}
               \displaystyle{\widehat{N}^{(\Psi)}+\frac{1}{2}\sum_{I=1}^4\theta_I^2
               \qquad\text{in the NS sector}~,}
               \\[3mm]
               \displaystyle{\widehat{N}^{(\Psi)}-\frac{1}{2}\sum_{I=1}^4\theta_I(1-\theta_I)
               \qquad\text{in the R sector}~,}
            \end{cases}
\label{L0Psi}
\end{equation}
where the fermionic number operator $\widehat{N}^{(\Psi)}$ is
\begin{equation}
\begin{aligned}
\widehat{N}^{(\Psi)}&=
\sum_{I=1}^4\bigg[\sum_{n=1-\nu}^\infty\!
(n-\theta_I) \overbar{\Psi}^{\dagger}_{I,\,n-\theta_I}\overbar{\Psi}_{n-\theta_I}^I
+\!\!\sum_{n=0+\nu}^\infty (n+\theta_I)\Psi^{\dagger\, I}_{n+\theta_I}
\Psi_{I,\,n+\theta_I}\bigg]
\\
&\quad\qquad+ \sum_{n=0+\nu}^\infty \!\!n\big(
\overbar{\Psi}^{\dagger}_{5,\,n}\overbar{\Psi}_{n}^5
+\Psi^{\dagger\,5}_{n}\Psi_{5,\,n}\big)~.
\end{aligned}
\label{Npsi}
\end{equation}
The $c$-numbers in (\ref{L0Z}) and (\ref{L0Psi}) arise from the normal ordering of 
the bosonic and fermionic oscillators.

The physical states must obey the conditions
\begin{equation}
\begin{cases}
               \displaystyle{L_0^{(Z)}+L_0^{(\Psi)}=\frac{1}{2}\qquad\text{in the NS sector}~,}
               \\[3mm]
               \displaystyle{L_0^{(Z)}+L_0^{(\Psi)}=0
               \qquad\text{in the R sector}~,}
            \end{cases}
\end{equation}
which imply Eq.~(\ref{phys}) of the main text.

\subsection*{BRST variations}
We now give a few details on how the BRST transformations of the instanton moduli
described in Section~\ref{secn:BRST} can be derived using the vertex operators introduced in
Section~\ref{secn:dinstantons}. To do so, we need the OPE's of the ten-dimensional string
theory, like for example
\begin{equation}
\begin{aligned}
S^{\dot{\mathcal{A}}}_{-1/2} (z)\,S^{\dot{\mathcal{B}}}_{-1/2} (w)\,&\sim\,\frac{\ii\,(\mathcal{C}\,
\Gamma_M)^{\dot{\mathcal{A}}\dot{\mathcal{B}}}}{\sqrt{2}}\,
\frac{\psi^M(w)\,\rme^{-\varphi(w)}}{z-w}+\ldots
\end{aligned}
\label{OPE10d}
\end{equation}
with $M=1,\ldots,10$.
Here $S^{\dot{\mathcal{A}}}_{-1/2}$ denotes a ten-dimensional spin field in the 
$(-\frac{1}{2})$-superghost picture with negative chirality. In particular,  
we have $S^{\dot{\mathcal{A}}}=S^A\,S^{\pm}\,\rme^{-\frac{1}{2}\,\varphi}$
where $S^A$ is the eight-dimensional spin field defined in (\ref{sfb}), with $+$ or $-$ depending
on whether $A$ is anti-chiral or chiral. The Dirac matrices $\Gamma^M$ and the charge conjugation matrix $\mathcal{C}$ are given in (\ref{10dgamma}) and (\ref{C10toC8}). Using these matrices
and the notation explained in Appendix~\ref{app:notations}, from (\ref{OPE10d}) we find
for instance
\begin{equation}
\big(SS^-\rme^{-\frac{1}{2}\,\varphi}\big)(z)~
\big(\overbar{S}_IS^+\rme^{-\frac{1}{2}\,\varphi}\big)(w)\,\sim\,-\frac{
\overbar{\Psi}_I(w)\,\rme^{-\varphi(w)}}{z-w}+\ldots~.
\label{OPE1}
\end{equation}

Let us now derive how the BRST charge (\ref{BRST}) acts on the moduli $B^I$ and $M^I$. As 
explained in the main text, we need to compute
\begin{align}
\Big[\mathcal{Q},V_{M^I}(w)\Big]=\frac{1}{\sqrt{2}}\,\Big[Q,V_{M^I}(w)\Big]+
\frac{1}{\sqrt{2}}\,\Big[\overbar{Q},V_{M^I}(w)\Big]~.
\end{align}
The second commutator vanishes and we are left with
\begin{equation}
\begin{aligned}
\Big[\mathcal{Q},V_{M^I}(w)\Big]&=\frac{1}{\sqrt{2}}\,\Big[Q,V_{M^I}(w)\Big]=-\frac{M^I}{\sqrt{2}}\,
\oint_w\frac{dz}{2\pi\ii}\,\big(SS^-\rme^{-\frac{1}{2}\,\varphi}\big)(z)
\,\big(\overbar{S}_IS^+\rme^{-\frac{1}{2}\,\varphi}\big)(w)
\end{aligned}
\end{equation}
where the minus sign is due to the fermionic statistics of $M^I$ and of the vertex operators. Using
the OPE (\ref{OPE1}), we immediately find
\begin{align}
\Big[\mathcal{Q},V_{M^I}(w)\Big]=\frac{M^I}{\sqrt{2}}\,
\overbar{\Psi}_I(w)\,\rme^{-\varphi(w)}~.
\end{align}
Since we have obtained the vertex of $B^I$ with a polarization $M^I$, we 
deduce that $\mathcal{Q}\,B^I=M^I$.

Proceeding systematically in this way, we obtain all BRST variations reported in the left columns
of (\ref{Q-1-1}) and (\ref{Qmixed}). The BRST variations in the right columns can be obtained by imposing the nilpotency of $\mathcal{Q}$ up to the symmetries of the system as explained in the
main text or, equivalently, by applying the vertex operator method in an enlarged system 
with auxiliary fields as discussed for example in \cite{Billo:2009di}.

\end{appendix}

\providecommand{\href}[2]{#2}\begingroup\raggedright\endgroup


\begin{thebibliography}{10}

\bibitem{Witten:1995im}
E.~Witten, {\it {Bound states of strings and p-branes}},  {\em Nucl. Phys. B}
  {\bf 460} (1996) 335--350, [\href{http://arxiv.org/abs/hep-th/9510135}{{\tt
  hep-th/9510135}}].

\bibitem{Douglas:1996uz}
M.~R. Douglas, {\it {Gauge fields and D-branes}},  {\em J. Geom. Phys.} {\bf
  28} (1998) 255--262, [\href{http://arxiv.org/abs/hep-th/9604198}{{\tt
  hep-th/9604198}}].

\bibitem{Green:1997tn}
M.~B. Green and M.~Gutperle, {\it {D-particle bound states and the D-instanton
  measure}},  {\em JHEP} {\bf 01} (1998) 005,
  [\href{http://arxiv.org/abs/hep-th/9711107}{{\tt hep-th/9711107}}].

\bibitem{Green:1998yf}
M.~B. Green and M.~Gutperle, {\it {D-instanton partition functions}},  {\em
  Phys. Rev.} {\bf D58} (1998) 046007,
  [\href{http://arxiv.org/abs/hep-th/9804123}{{\tt hep-th/9804123}}].

\bibitem{Green:2000ke}
M.~B. Green and M.~Gutperle, {\it {D-instanton induced interactions on a
  D3-brane}},  {\em JHEP} {\bf 02} (2000) 014,
  [\href{http://arxiv.org/abs/hep-th/0002011}{{\tt hep-th/0002011}}].

\bibitem{Billo:2002hm}
M.~Billo, M.~Frau, I.~Pesando, F.~Fucito, A.~Lerda, and A.~Liccardo, {\it
  {Classical gauge instantons from open strings}},  {\em JHEP} {\bf 02} (2003)
  045, [\href{http://arxiv.org/abs/hep-th/0211250}{{\tt hep-th/0211250}}].

\bibitem{Maccaferri:2018vwo}
C.~Maccaferri and A.~Merlano, {\it {Localization of effective actions in open
  superstring field theory}},  {\em JHEP} {\bf 03} (2018) 112,
  [\href{http://arxiv.org/abs/1801.07607}{{\tt arXiv:1801.07607}}].

\bibitem{Blumenhagen:2006xt}
R.~Blumenhagen, M.~Cvetic, and T.~Weigand, {\it {Spacetime instanton
  corrections in 4D string vacua - the seesaw mechanism for D-brane models}},
  {\em Nucl. Phys.} {\bf B771} (2007) 113--142,
  [\href{http://arxiv.org/abs/hep-th/0609191}{{\tt hep-th/0609191}}].

\bibitem{Ibanez:2006da}
L.~E. Ibanez and A.~M. Uranga, {\it {Neutrino Majorana masses from string
  theory instanton effects}},  {\em JHEP} {\bf 03} (2007) 052,
  [\href{http://arxiv.org/abs/hep-th/0609213}{{\tt hep-th/0609213}}].

\bibitem{Argurio:2006ny}
R.~Argurio, M.~Bertolini, S.~Franco, and S.~Kachru, {\it {Gauge/gravity duality
  and meta-stable dynamical supersymmetry breaking}},  {\em JHEP} {\bf 01}
  (2007) 083, [\href{http://arxiv.org/abs/hep-th/0610212}{{\tt
  hep-th/0610212}}].

\bibitem{Argurio:2007vqa}
R.~Argurio, M.~Bertolini, G.~Ferretti, A.~Lerda, and C.~Petersson, {\it
  {Stringy Instantons at Orbifold Singularities}},  {\em JHEP} {\bf 06} (2007)
  067, [\href{http://arxiv.org/abs/0704.0262}{{\tt arXiv:0704.0262}}].

\bibitem{Bianchi:2007wy}
M.~Bianchi, F.~Fucito, and J.~F. Morales, {\it {D-brane Instantons on the
  orientifold}},  {\em JHEP} {\bf 07} (2007) 038,
  [\href{http://arxiv.org/abs/0704.0784}{{\tt arXiv:0704.0784}}].

\bibitem{Blumenhagen:2007zk}
R.~Blumenhagen, M.~Cvetic, D.~Lust, R.~Richter, and T.~Weigand, {\it
  {Non-perturbative Yukawa Couplings from String Instantons}},  {\em Phys. Rev.
  Lett.} {\bf 100} (2008) 061602, [\href{http://arxiv.org/abs/0707.1871}{{\tt
  arXiv:0707.1871}}].

\bibitem{Ibanez:2007rs}
L.~E. Ibanez, A.~N. Schellekens, and A.~M. Uranga, {\it {Instanton Induced
  Neutrino Majorana Masses in CFT Orientifolds with MSSM-like spectra}},  {\em
  JHEP} {\bf 06} (2007) 011, [\href{http://arxiv.org/abs/0704.1079}{{\tt
  arXiv:0704.1079}}].

\bibitem{Ibanez:2007tu}
L.~E. Ibanez and A.~M. Uranga, {\it {Instanton Induced Open String
  Superpotentials and Branes at Singularities}},  {\em JHEP} {\bf 02} (2008)
  103, [\href{http://arxiv.org/abs/0711.1316}{{\tt arXiv:0711.1316}}].

\bibitem{Billo:2009di}
M.~Billo, L.~Ferro, M.~Frau, L.~Gallot, A.~Lerda, and I.~Pesando, {\it {Exotic
  instanton counting and heterotic/type I' duality}},  {\em JHEP} {\bf 07}
  (2009) 092, [\href{http://arxiv.org/abs/0905.4586}{{\tt arXiv:0905.4586}}].

\bibitem{Fucito:2009rs}
F.~Fucito, J.~F. Morales, and R.~Poghossian, {\it {Exotic prepotentials from
  D(-1)D7 dynamics}},  {\em JHEP} {\bf 10} (2009) 041,
  [\href{http://arxiv.org/abs/0906.3802}{{\tt arXiv:0906.3802}}].

\bibitem{Polchinski:1995df}
J.~Polchinski and E.~Witten, {\it {Evidence for Heterotic - Type I String
  Duality}},  {\em Nucl. Phys.} {\bf B460} (1996) 525--540,
  [\href{http://arxiv.org/abs/hep-th/9510169}{{\tt hep-th/9510169}}].

\bibitem{Witten:2000mf}
E.~Witten, {\it {BPS Bound states of D0 - D6 and D0 - D8 systems in a B
  field}},  {\em JHEP} {\bf 04} (2002) 012,
  [\href{http://arxiv.org/abs/hep-th/0012054}{{\tt hep-th/0012054}}].

\bibitem{Nekrasov:2002qd}
N.~Nekrasov, {\it {Seiberg-Witten prepotential from instanton counting}},  {\em
  Adv. Theor. Math. Phys.} {\bf 7} (2004) 831--864,
  [\href{http://arxiv.org/abs/hep-th/0206161}{{\tt hep-th/0206161}}].

\bibitem{Moore:1998et}
G.~W. Moore, N.~Nekrasov, and S.~Shatashvili, {\it {D-particle bound states and
  generalized instantons}},  {\em Commun. Math. Phys.} {\bf 209} (2000) 77--95,
  [\href{http://arxiv.org/abs/hep-th/9803265}{{\tt hep-th/9803265}}].

\bibitem{Flume:2002az}
R.~Flume and R.~Poghossian, {\it {An algorithm for the microscopic evaluation
  of the coefficients of the Seiberg-Witten prepotential}},  {\em Int. J. Mod.
  Phys.} {\bf A18} (2003) 2541,
  [\href{http://arxiv.org/abs/hep-th/0208176}{{\tt hep-th/0208176}}].

\bibitem{Bruzzo:2002xf}
U.~Bruzzo, F.~Fucito, J.~F. Morales, and A.~Tanzini, {\it {Multi-instanton
  calculus and equivariant cohomology}},  {\em JHEP} {\bf 05} (2003) 054,
  [\href{http://arxiv.org/abs/hep-th/0211108}{{\tt hep-th/0211108}}].

\bibitem{Nekrasov:2003rj}
N.~Nekrasov and A.~Okounkov, {\it {Seiberg-Witten theory and random
  partitions}},  {\em Prog. Math.} {\bf 244} (2006) 525--596,
  [\href{http://arxiv.org/abs/hep-th/0306238}{{\tt hep-th/0306238}}].

\bibitem{Bruzzo:2003rw}
U.~Bruzzo and F.~Fucito, {\it {Superlocalization formulas and supersymmetric
  Yang-Mills theories}},  {\em Nucl. Phys.} {\bf B678} (2004) 638--655,
  [\href{http://arxiv.org/abs/math-ph/0310036}{{\tt math-ph/0310036}}].

\bibitem{Marino:2004cn}
M.~Marino and N.~Wyllard, {\it {A note on instanton counting for N = 2 gauge
  theories with classical gauge groups}},  {\em JHEP} {\bf 05} (2004) 021,
  [\href{http://arxiv.org/abs/hep-th/0404125}{{\tt hep-th/0404125}}].

\bibitem{Nekrasov:2017cih}
N.~Nekrasov, {\it {Magnificent Four}},
  \href{http://arxiv.org/abs/1712.08128}{{\tt arXiv:1712.08128}}.

\bibitem{Cao:2017swr}
Y.~Cao and M.~Kool, {\it {Zero-dimensional Donaldson\textendash{}Thomas
  invariants of Calabi\textendash{}Yau 4-folds}},  {\em Adv. Math.} {\bf 338}
  (2018) 601--648, [\href{http://arxiv.org/abs/1712.07347}{{\tt
  arXiv:1712.07347}}].

\bibitem{Nekrasov:2018xsb}
N.~Nekrasov and N.~Piazzalunga, {\it {Magnificent Four with Colors}},  {\em
  Commun. Math. Phys.} {\bf 372} (2019), no.~2 573--597,
  [\href{http://arxiv.org/abs/1808.05206}{{\tt arXiv:1808.05206}}].

\bibitem{Fucito:2020bjd}
F.~Fucito, J.~Morales, and R.~Poghossian, {\it {The Chiral Ring of $N=2$ in
  Eight Dimensions}},  \href{http://arxiv.org/abs/2010.10235}{{\tt
  arXiv:2010.10235}}.

\bibitem{Billo:2006jm}
M.~Billo, M.~Frau, F.~Fucito, and A.~Lerda, {\it {Instanton calculus in R-R
  background and the topological string}},  {\em JHEP} {\bf 11} (2006) 012,
  [\href{http://arxiv.org/abs/hep-th/0606013}{{\tt hep-th/0606013}}].

\bibitem{Bonelli:2020gku}
G.~Bonelli, N.~Fasola, A.~Tanzini, and Y.~Zenkevich, {\it {ADHM in 8d, coloured
  solid partitions and Donaldson-Thomas invariants on orbifolds}},
  \href{http://arxiv.org/abs/2011.02366}{{\tt arXiv:2011.02366}}.

\bibitem{Tseytlin:1986ti}
A.~A. Tseytlin, {\it {Vector Field Effective Action in the Open Superstring
  Theory}},  {\em Nucl. Phys. B} {\bf 276} (1986) 391. [Erratum: Nucl.Phys.B
  291, 876 (1987)].

\bibitem{Billo':2009gc}
M.~Billo, M.~Frau, L.~Gallot, A.~Lerda, and I.~Pesando, {\it {Classical
  solutions for exotic instantons?}},  {\em JHEP} {\bf 03} (2009) 056,
  [\href{http://arxiv.org/abs/0901.1666}{{\tt arXiv:0901.1666}}].
  
\bibitem{Bertolini:2005qh}
M.~Bertolini, M.~Billo, A.~Lerda, J.~F. Morales, and R.~Russo, {\it {Brane
  world effective actions for D-branes with fluxes}},  {\em Nucl. Phys.} {\bf
  B743} (2006) 1--40, [\href{http://arxiv.org/abs/hep-th/0512067}{{\tt
  hep-th/0512067}}].

\bibitem{Dixon:1986qv}
L.~J. Dixon, D.~Friedan, E.~J. Martinec, and S.~H. Shenker, {\it {The Conformal
  Field Theory of Orbifolds}},  {\em Nucl. Phys.} {\bf B282} (1987) 13--73.

\bibitem{Friedan:1985ge}
D.~Friedan, E.~J. Martinec, and S.~H. Shenker, {\it {Conformal Invariance,
  Supersymmetry and String Theory}},  {\em Nucl. Phys.} {\bf B271} (1986) 93.

\bibitem{Billo:1998vr}
M.~Billo, P.~Di~Vecchia, M.~Frau, A.~Lerda, I.~Pesando, R.~Russo, and
  S.~Sciuto, {\it {Microscopic string analysis of the D0-D8 brane system and
  dual R-R states}},  {\em Nucl. Phys.} {\bf B526} (1998) 199--228,
  [\href{http://arxiv.org/abs/hep-th/9802088}{{\tt hep-th/9802088}}].

\bibitem{Bergman:1997rf}
O.~Bergman and M.~R. Gaberdiel, {\it {A Nonsupersymmetric open string theory
  and S duality}},  {\em Nucl. Phys. B} {\bf 499} (1997) 183--204,
  [\href{http://arxiv.org/abs/hep-th/9701137}{{\tt hep-th/9701137}}].

\bibitem{Bergman:1997gf}
O.~Bergman, M.~R. Gaberdiel, and G.~Lifschytz, {\it {Branes, orientifolds and
  the creation of elementary strings}},  {\em Nucl. Phys. B} {\bf 509} (1998)
  194--215, [\href{http://arxiv.org/abs/hep-th/9705130}{{\tt hep-th/9705130}}].

\bibitem{Green:1987mn}
M.~B. Green, J.~Schwarz, and E.~Witten, {\em {Superstring Theory. Vol. 2: Loop
  amplitudes, Anomalies and Phenomenology}}.
\newblock Cambridge University Press, 1988.

\bibitem{Kitao:1998vn}
T.~Kitao, N.~Ohta, and J.-G. Zhou, {\it {Fermionic zero mode and string
  creation between D4-branes at angles}},  {\em Phys. Lett. B} {\bf 428} (1998)
  68--74, [\href{http://arxiv.org/abs/hep-th/9801135}{{\tt hep-th/9801135}}].

\bibitem{Sen:2020cef}
A.~Sen, {\it {D-instanton Perturbation Theory}},  {\em JHEP} {\bf 08} (2020)
  075, [\href{http://arxiv.org/abs/2002.04043}{{\tt arXiv:2002.04043}}].

\bibitem{Billo:2004zq}
M.~Billo, M.~Frau, I.~Pesando, and A.~Lerda, {\it {N = 1/2 gauge theory and its
  instanton moduli space from open strings in R-R background}},  {\em JHEP}
  {\bf 05} (2004) 023, [\href{http://arxiv.org/abs/hep-th/0402160}{{\tt
  hep-th/0402160}}].

\bibitem{JK1995}
L.~C. Jeffrey and F.~C. Kirwan, {\it {Surface Operators and Separation of
  Variables}},  {\em Topology} {\bf 34} (1995) 291--327.

\bibitem{Grossman:1984pi}
B.~Grossman, T.~W. Kephart, and J.~D. Stasheff, {\it {Solutions to Yang-Mills
  field equations in eight-dimensions and the last Hopf map}},  {\em Commun.
  Math. Phys.} {\bf 96} (1984) 431 (Erratum: ibid. {\bf 100} (1985) 311).

\bibitem{Grossman:1989bb}
B.~Grossman, T.~W. Kephart, and J.~D. Stasheff, {\it {Solutions to gauge field
  equations in eight-dimensions: conformal invariance and the last Hopf map}},
  {\em Phys. Lett.} {\bf B220} (1989) 431.
  
\end{thebibliography}
\end{document}